\documentclass[floats,floatfix,showpacs,amssymb,prd,superscriptaddress,nofootinbib,nolongbibliography,reprint]{revtex4-1}

\usepackage{amssymb,amsmath,verbatim,mathtools,needspace,enumitem,etoolbox,graphicx,physics,microtype,afterpage,xspace,tabularx,lmodern,multirow,bm}
\usepackage{textcomp, gensymb}
\usepackage{float}
\usepackage{braket}

\usepackage[table]{xcolor}
\usepackage{color}
\usepackage{colortbl}
\usepackage{multirow}
\usepackage{tikz,lipsum,lmodern}
\usepackage[most]{tcolorbox}

\usepackage{relsize}

\usepackage{dcolumn}

\usepackage[normalem]{ulem}
\usepackage{placeins}
\usepackage{comment}
\definecolor{linkcolor}{rgb}{0.0,0.3,0.5}
\usepackage[unicode, colorlinks=true, linkcolor=linkcolor, citecolor=linkcolor, filecolor=linkcolor, urlcolor=linkcolor, linktocpage, breaklinks]{hyperref}
\usepackage[all]{hypcap}
\usepackage[T1]{fontenc}
\usepackage[utf8]{inputenc}
\usepackage{mathrsfs}
\usepackage{breqn}
\hypersetup{colorlinks=true,citecolor=romared,linkcolor=romared,urlcolor=romared}

\definecolor{romared}{RGB}{142,0,28}

\newcommand{\be}{\begin{equation}}
\newcommand{\ee}{\end{equation}}

\def\be{\begin{equation}}
\def\ee{\end{equation}}
\newcommand{\beq}{\begin{eqnarray}}
\newcommand{\eeq}{\end{eqnarray}}
\usepackage{makecell}
\usepackage{soul}
\usepackage[nolist,nohyperlinks]{acronym}
\usepackage{lipsum}
\usepackage{braket}

\newcommand{\jhu}{William H. Miller III Department of Physics and Astronomy, Johns Hopkins University, 3400 North Charles Street, Baltimore, Maryland, 21218, USA}

\graphicspath{{../Plots/}}

\newcommand{\ben}{\begin{enumerate}}
\newcommand{\een}{\end{enumerate}}

\def\be{\begin{equation}}
\def\ee{\end{equation}}
\def\beq{\begin{eqnarray}}
\def\eeq{\end{eqnarray}}

\newcommand{\sapienza}{Dipartimento di Fisica, Sapienza Università 
	di Roma, Piazzale Aldo Moro 5, 00185, Roma, Italy}
\newcommand{\infn}{INFN, Sezione di Roma, Piazzale Aldo Moro 2, 00185, Roma, Italy}

\graphicspath{{./Plots/}}

\allowdisplaybreaks

\begin{document}

\pagenumbering{arabic}

\title{Modeling the frequency-domain ringdown amplitude\\ of comparable-mass mergers with greybody factors
}

\author{Romeo Felice Rosato}
\email{romeofelice.rosato@uniroma1.it}
\affiliation{\sapienza}
\affiliation{\infn}

\author{Sophia Yi}
\email{syi24@jh.edu}
\affiliation{\jhu}

\author{Emanuele Berti}
\email{berti@jhu.edu}
\affiliation{\jhu}

\author{Paolo Pani}
\email{paolo.pani@uniroma1.it}
\affiliation{\sapienza}
\affiliation{\infn}

\pacs{}
\date{\today}

\begin{abstract}
It was recently shown that, in a binary coalescence, the greybody factor of the remnant black hole modulates the post-merger ringdown signal. In this work, we demonstrate that a simple four-parameter model based on the greybody factor accurately reproduces the frequency-domain amplitude of a large set of comparable-mass, aligned-spin numerical relativity waveforms from the SXS catalog, achieving mismatches of order ${\cal O}(10^{-5})$ and improving existing models by roughly two orders of magnitude. We also identify the optimal initial frequency for applying the model in the frequency domain and provide analytical fits of the model parameters in terms of the progenitor masses and aligned spins. Our results pave the way for new consistency tests of the ringdown phase, complementary to traditional black hole spectroscopy.
\end{abstract}

\maketitle

\section{Introduction}
The final stage of a binary black hole (BBH) coalescence has long been regarded as one of the most promising regimes for testing strong-field gravity. In this ``ringdown’’ phase, the gravitational waveform is customarily decomposed into the decaying oscillation modes of the remnant black hole (BH) (see~\cite{Berti:2025hly} for a recent review). Within general relativity~(GR), these quasinormal modes (QNMs) have complex frequencies determined uniquely by the mass and angular momentum of the final (Kerr) BH. This property underlies a broad class of tests of GR in which multiple QNMs are extracted from the data and checked for mutual consistency with the Kerr prediction.

In practice, however, accurately measuring QNM parameters is notoriously challenging. Even in the loudest event observed to date, GW250114~\cite{LIGOScientific:2025rid,LIGOScientific:2025obp} the LIGO-Virgo-KAGRA (LVK) Collaboration has only recently achieved confident measurements of two QNMs, showing an agreement with the GR prediction at the level of $\approx 30\%$. A major difficulty arises from uncertainties in selecting the starting time of the ringdown analysis and in choosing the appropriate number of QNMs to include in the template. Although BH perturbation theory predicts an infinite tower of QNMs, only a few are expected to be observable in realistic BBH mergers. Without reliable guidance on which modes are excited and detectable at a given time~\cite{Crescimbeni:2025ytx}, QNM analyses risk overfitting the data with an excessive number of free parameters. These issues have led to substantial debate regarding the robustness of QNM measurements in several BBH events~\cite{Carullo:2019flw,Isi:2019aib,Cotesta:2022pci,Finch:2022ynt,Ma:2023vvr,Isi:2023nif,Carullo:2023gtf,Capano:2021etf,Siegel:2023lxl}.

Because only a small subset of QNMs are measurable, an independent determination of the mass~$M$ and dimensionless spin~$\chi=J/M^2$ of the remnant BH is highly desirable (throughout this paper, we use $G=c=1$ units). Current consistency tests infer $(M,\chi)$ either from the early portion of the signal (inspiral and early merger) or from the full waveform~\cite{LIGOScientific:2020tif,LIGOScientific:2021sio}, and compare these estimates with those obtained from a handful of QNMs. This strategy, however, inherits any systematic uncertainties present in the inspiral and merger modeling, which may bias the inferred remnant properties~\cite{Gupta:2024gun} and weaken the overall test of GR.

There are various approaches based on numerical-relativity surrogates and post-peak analyses that can help to mitigate such sources of uncertainty~\cite{CalderonBustillo:2020rmh,Chandra:2025ipu,Gennari:2023gmx,Kankani:2025gqj}. Nevertheless, these challenges motivate the search for a complementary, non-QNM-centric description of the ringdown that could both improve our physical understanding of the remnant, provide an independent way to estimate~$(M,\chi)$, and
allow for model-agnostic tests of gravity in the ringdown. This search has led to recent increased interest in the BH greybody factor, which quantifies the absorptive nature of the BH geometry. Specifically, the greybody factor is given by the transmissivity of the BH effective potential, and since it depends only on the geometry of the BH, it potentially offers a complementary way to extract the remnant mass and spin.

It was recently suggested that the greybody factor of the remnant BH
hole modulates the ringdown amplitude in the frequency domain~\cite{Oshita:2023cjz}, offering a complementary way to extract the remnant mass and spin.
Although originally developed within BH perturbation theory for the signal emitted by a test particle plunging into a BH~\cite{Oshita:2023cjz,Rosato:2024arw,Rosato:2025byu}, this model was later shown to describe remarkably well the ringdown of comparable-mass mergers~\cite{Okabayashi:2024qbz}. Moreover, greybody factors are known to be stable under small perturbations of the effective potential --~such as those induced by environmental or beyond-GR effects~\cite{Rosato:2024arw,Oshita:2024fzf,Rosato:2025lxb}~-- in sharp contrast with the well-studied spectral instability of QNMs~\cite{Nollert:1996rf,Barausse:2014tra,Jaramillo:2020tuu,Cheung:2021bol,Berti:2022xfj}.

In this work, we investigate how broadly the greybody-factor description applies to comparable-mass BBH mergers, and assess its potential as an alternative parameterization of the ringdown. Our goal is to determine whether this framework can ultimately enable independent measurements of remnant masses and spins, thereby providing a complementary avenue for testing GR beyond traditional QNM analyses.

For this purpose, in Sec.~\ref{sec:model} we introduce a simple phenomenological model for the ringdown amplitude in the frequency domain, inspired by the parameterization proposed in~\cite{Rosato:2024arw,Rosato:2025lxb}. The waveform is proportional to the BH greybody factor, and thus depends only on the remnant mass~$M$ and spin~$\chi$, and contains two additional parameters. One of these parameters controls an extra power-law frequency modulation relative to the greybody factor~\cite{Rosato:2024arw,Rosato:2025lxb}, whose inclusion improves the accuracy of the model by roughly two orders of magnitude.

We validate this model by fitting it to a large set of numerical relativity simulations of comparable-mass, aligned-spin BBH mergers from the SXS catalog~\cite{Boyle:2019kee,Scheel:2025jct}.
In Sec.~\ref{sec:fits}, we analyze how the model parameters depend on the progenitor masses and (aligned) spins, and we provide practical fitting formulas suitable for parameter-estimation applications.
We then test the predictive accuracy of the fitted model on a large set of simulations.
Finally, in Sec.~\ref{sec:discussion} we summarize our main results and discuss future prospects.

\section{Modeling the gravitational-wave spectrum in the frequency domain} \label{sec:model}
In this section, we introduce our model of the frequency-domain spectral amplitude 
of the gravitational-wave signal, $H_{\ell m}(\omega)$, with $\ell$ and $m$ 
indicating the angular-mode spherical harmonic indices (we neglect spherical-spheroidal mode mixing: see Appendix~\ref{app:greybody_teukolsky}). The model expresses $H_{\ell m}(\omega)$ in terms of the BH reflectivity 
$\mathcal{R}_{\ell m}(M\omega,\chi)$, using numerical relativity simulations of BBH coalescences from the SXS catalog~\cite{Boyle:2019kee,Scheel:2025jct}. For brevity, we introduce the dimensionless quantity $\bar{\omega}\equiv M\omega$.

The quantity $H_{\ell m}(\omega)$ is defined as the absolute value of the Fourier transform of the time-domain signal, i.e. $H_{\ell m}(\omega)=|h_{\ell m}(\omega)|$, adopting the convention
\begin{equation}
    h_{\ell m}(\omega) = \int dt\, e^{+i\omega t}\, h_{\ell m}(t)\,,
\end{equation}
with $h_{\ell m}(t)$ being the full complex time-domain signal.
Numerical subtleties related to its evaluation from SXS data, and the associated frequency range where the spectrum is reliable, are discussed in Appendix~\ref{app:fft}.

The BH reflectivity $\mathcal{R}_{\ell m}(\bar{\omega},\chi)$ 
encodes the reflection probability associated with the gravitational effective potential. 
It is related to the transmission probability $\Gamma_{\ell m}(\bar{\omega},\chi)$ (i.e., the greybody factor) through 
\begin{equation}\label{eq:gfsrefl}
\mathcal{R}_{\ell m}(\bar{\omega},\chi)=1-{\omega-m\Omega_{\rm H} \over\omega}\Gamma_{\ell m}(\bar{\omega},\chi)\,,
\end{equation}
where $\Omega_{\rm H} $ is the horizon angular velocity.
This quantity can be computed by directly integrating the Teukolsky equation with plane-wave scattering boundary conditions~\cite{Teukolsky:1972my,Press:1972zz,Press:1972zz,Starobinskii:1973vzb,Starobinskil:1974nkd} (see Appendix~\ref{app:greybody_teukolsky} for a detailed discussion). For a Kerr BH, it is uniquely determined by the remnant mass $M$ and dimensionless spin $\chi$. Several recent works have shown that this reflectivity leaves a distinct imprint on the 
frequency-domain spectrum of the ringdown signal, 
both in the point-particle limit~\cite{Oshita:2023cjz,Rosato:2024arw,Rosato:2025byu} and for comparable mass mergers~\cite{Okabayashi:2024qbz}.
These findings motivate the construction of spectral models directly based on $\mathcal{R}_{\ell m}(\bar{\omega},\chi)$,
bridging numerical relativity results with the underlying BH scattering properties.
Our goal here is to refine and validate such a model, and to assess its performance against 
previous formulations proposed in the literature.
\subsection{Validation of the reflectivity-based spectral model}\label{sec:modelcomparison}
Previous analyses have modeled the frequency-domain amplitude using 
$H^{(1)}_{\ell m} = M A_{\ell m}\,\sqrt{\mathcal{R}_{\ell m}}$~\cite{Okabayashi:2024qbz}, were $A_{\ell m}$ is the overall amplitude of the $(\ell,m)$ mode.
Here we adopt a modified model, motivated by the plunging-particle theoretical scenario developed in Refs.~\cite{Rosato:2024arw,Oshita:2024fzf,Rosato:2025byu}:
\begin{equation}\label{eq:model}
H^{(2)}_{\ell m}(\omega)=M\frac{A_{\ell m}\,\sqrt{\mathcal{R}_{\ell m}(\bar{\omega},\chi)}}{\bar{\omega}^{p_{\ell m}}}\,,
\end{equation}
where $p_{\ell m}$ is an extra parameter that introduces an additional, and generally source-dependent, frequency modulation. As we shall show, the model above provides a significantly improved description of the numerical data.
In Eq.~\eqref{eq:model}, $M$ and $\chi$ denote the remnant mass and spin, while $A_{\ell m}$ and $p_{\ell m}$ are dimensionless model parameters depending on the multipole under consideration.

Our first goal is to validate the modified model introduced in Eq.~\eqref{eq:model}
and to compare its performance against previous formulations,
as a first step toward motivating the subsequent analysis and results presented throughout this work.
\begin{figure}[ht]
    \centering
    \includegraphics[width=\linewidth]{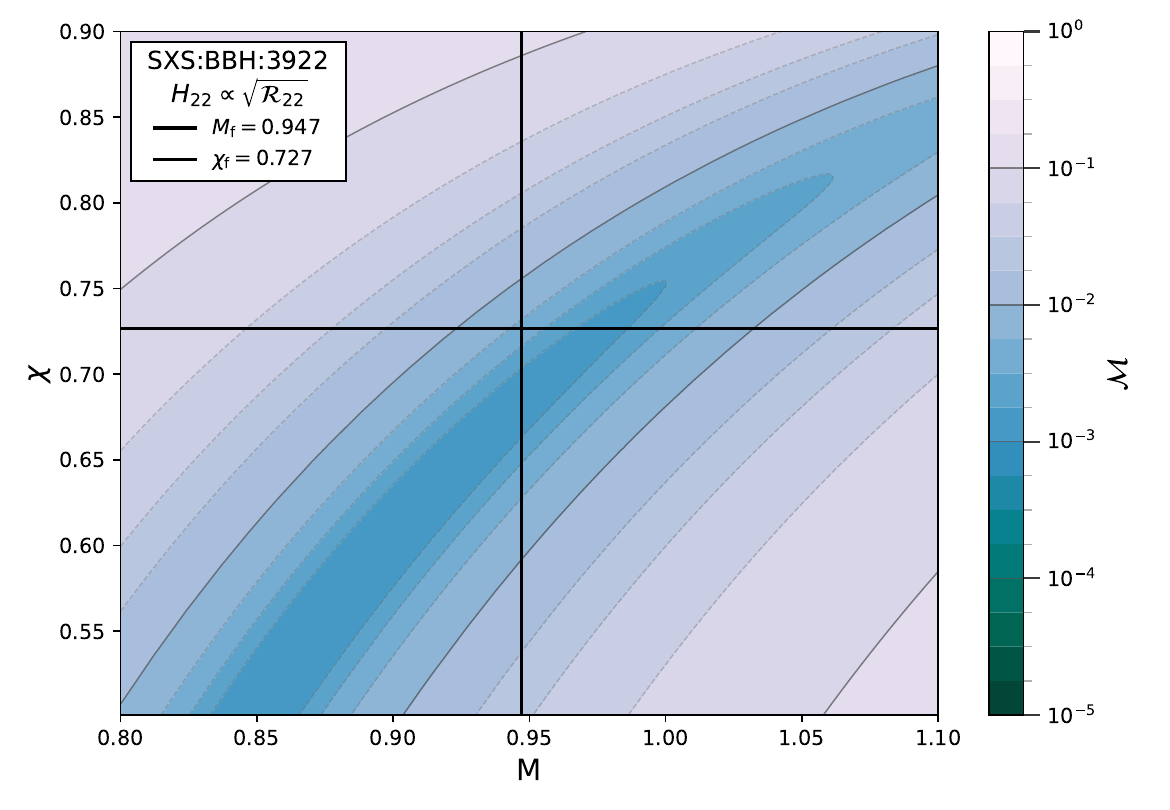}
    \includegraphics[width=\linewidth]{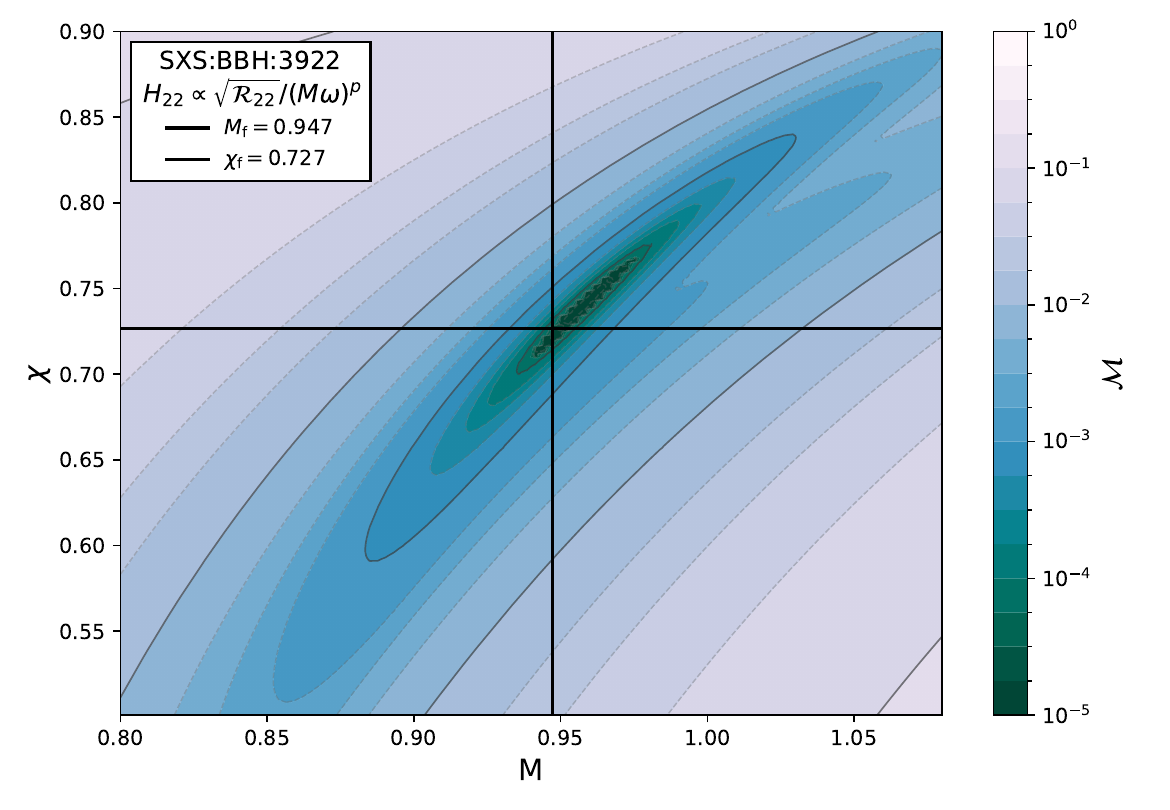}
   \caption{
Mismatch $\mathcal{M}$, defined in Eq.~\eqref{eq:mismatch}, between the numerical Fourier transform  
$H_{\ell m}^{\rm data}(\omega) = |h_{\ell m}(\omega)|$ 
from the simulation \texttt{SXS:BBH:3922} and two analytical models.
The mismatch is shown over a portion of the $(M,\chi)$ plane of the remnant.
The upper panel corresponds to the model 
$H^{(1)}_{\ell m}(\omega) = MA_{\ell m}\,\sqrt{\mathcal{R}_{\ell m}(\bar{\omega},\chi)}$ proposed in Ref.~\cite{Okabayashi:2024qbz},
for which the fit is performed with a single free parameter $A_{\ell m}$,
while the lower panel shows results for the modified model proposed in this work,
$H^{(2)}_{\ell m}(\omega) = MA_{\ell m}\,\sqrt{\mathcal{R}_{\ell m}(\bar{\omega},\chi)}/\bar{\omega}^{p_{\ell m}}$ (see Eq.~\eqref{eq:model}),
where both $A_{\ell m}$ and $p_{\ell m}$ are fitted. The fit is performed over the range $\bar{\omega} \in [0.35,\,0.75]$, 
determined for this specific simulation as discussed in Appendix~\ref{subsec:fittingprocedure}.
For the model $H^{(1)}_{\ell m}$, the region of minimal mismatch does not exactly coincide with the true remnant parameters,
whereas for the modified model $H^{(2)}_{\ell m}$ the true values lie within the minimum-mismatch region.
Overall, the agreement improves by about two orders of magnitude, 
with the mismatch $\mathcal{M}$ dropping from $\sim\!10^{-3}$ to $\sim\!10^{-5}$ 
around the true values of the remnant parameters.
}
    \label{fig:modelcomparison}
\end{figure}
Figure~\ref{fig:modelcomparison} illustrates why the model
$H^{(2)}_{\ell m}$
captures the data more accurately than the simpler
$H^{(1)}_{\ell m}$.
We show results for the SXS simulation \texttt{SXS:BBH:3922},
but similar behavior is observed across the entire catalog.
The figure reports the mismatch between the numerical spectrum and each model
in the $(M,\chi)$ plane. The mismatch $\mathcal{M}$ is defined as
\begin{equation}\label{eq:mismatch}
    \mathcal{M}= 1-{\braket{H^{\rm data} |H^{\rm model}} \over \sqrt{\braket{H^{\rm data}|H^{\rm data}} \braket{H^{\rm model}|H^{\rm model}}}}\,,
\end{equation}
where the bracket denotes the scalar product between two functions of $\bar{\omega}$, defined as
\begin{equation}
    \langle a\mid b\rangle
    =
    \int_{\bar{\omega}_i}^{\bar{\omega}_f} \! d\bar{\omega} \, a^*(\bar{\omega})\, b(\bar{\omega})\, ,
\end{equation}
$\bar{\omega}_i$ and $\bar{\omega}_f$ define the boundaries of the fitting domain, and the complex conjugation is trivial for the real-valued functions considered here.

The upper panel of Fig.~\ref{fig:modelcomparison} shows the results obtained with model $H^{(1)}$. Although the mismatch reaches values of order $\mathcal{M}\sim10^{-3}$, the minimum-mismatch region in the $(M,\chi)$ plane is relatively broad, indicating a significant degeneracy between the remnant mass and spin, and therefore a weak constraint on the remnant parameters.
In contrast, the lower panel shows that for model $H^{(2)}$ the mismatch improves by about two orders of magnitude,
reaching values of order $\mathcal{M}\sim10^{-5}$, and that the region of minimal mismatch
is substantially more localized.
This leads to a significantly tighter constraint on the remnant parameters,
with a strong reduction of the $(M,\chi)$ degeneracy, and with the true values of these parameters lying well within the minimum-mismatch region.

\subsection{Fitting procedure}\label{subsec:fittingprocedure}

Given the model's preliminarily good agreement with some numerical relativity data, we have tested it on a large number of simulations from the SXS catalog to assess its performance across different physical configurations. In this section, we describe how to reliably extract $A_{\ell m}$ and $p_{\ell m}$ from a single simulation.

\begin{figure*}[ht]
    \centering\includegraphics[width=\linewidth]{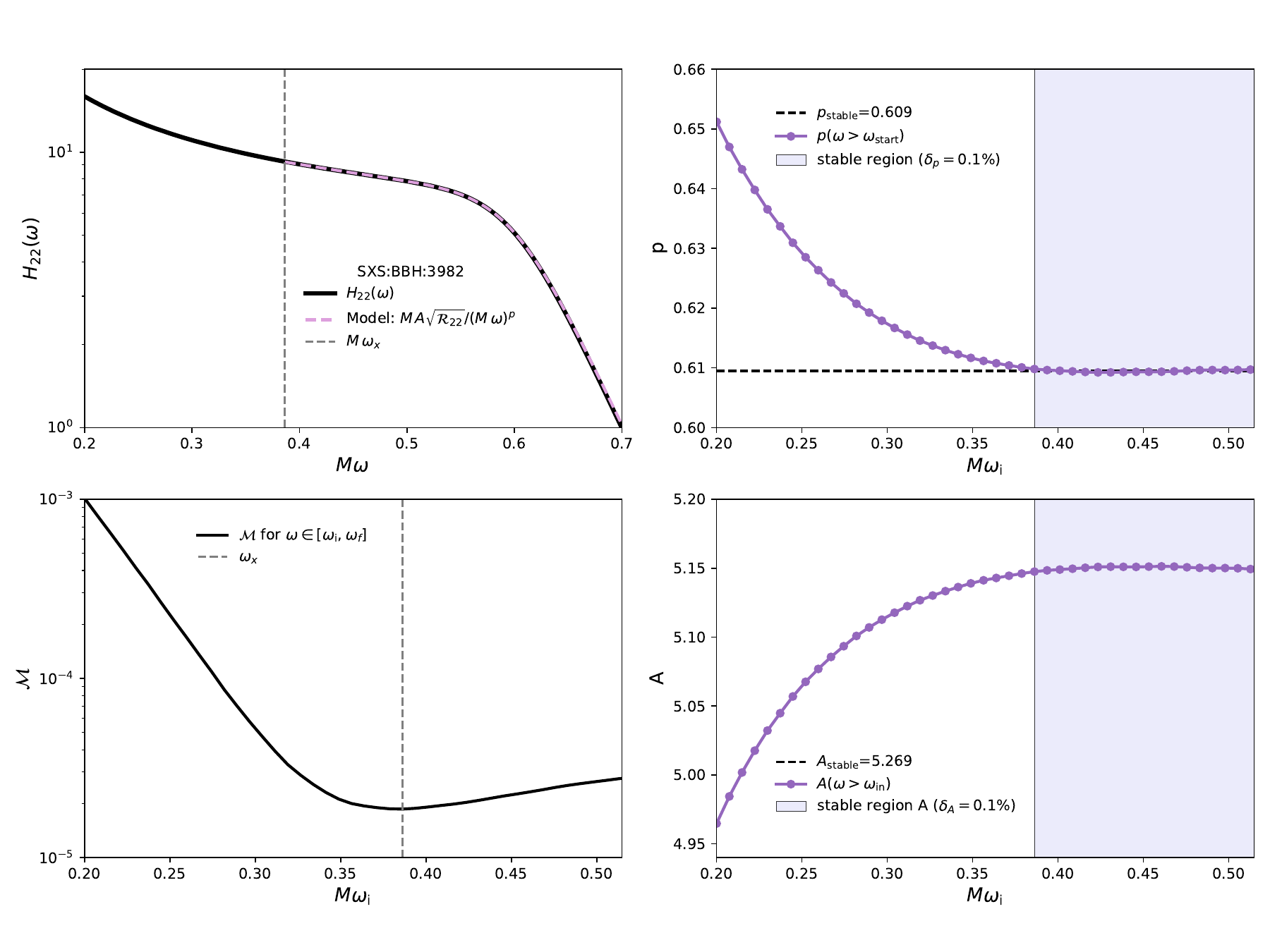}
    \caption{
Illustration of the fitting procedure for the simulation \texttt{SXS:BBH:3982},
considering the dominant $(\ell,m)=(2,2)$ mode.
The goal of the fit is both to extract the model parameters $A_{\ell m}$ and $p_{\ell m}$ 
and to determine the frequency range over which the model remains valid.
For this purpose, we perform fits over intervals $\bar{\omega} \in [\bar{\omega}_i,\,\bar{\omega}_{\rm cut}]$ 
for several starting frequencies $\bar{\omega}_i$.
The top-left panel shows the numerical frequency-domain spectrum 
$H_{22}(\omega)=|h_{22}(\omega)|$ (solid black) 
and the best-fit model $A/\bar{\omega}^p$ (dashed pink),
evaluated for $\bar{\omega} < \bar{\omega}_{\rm cut}$ as described in Appendix~\ref{app:fft}.
The bottom-left panel displays the mismatch $\mathcal{M}$, computed according to Eq.~\eqref{eq:mismatch},
as a function of the initial fitting frequency $\bar{\omega}_i$.
The top-right and bottom-right panels show the fitted parameters $p_{\ell m}$ and $A_{\ell m}$ as functions of $\bar{\omega}_i$.
Both parameters become stable beyond a certain frequency $\bar{\omega}_x$, 
indicated by the shaded region, 
where their variations remain below $0.1\%$.
The frequency $\bar{\omega}_x$ therefore marks the lower bound of validity of the model.
The mismatch reaches its minimum in this stability region,
with $\mathcal{M}\simeq2\cdot10^{-5}$,
demonstrating the excellent agreement between the model and the numerical data.
}
\label{fig:fittingprocedure}
\end{figure*}

An example of the Fourier-transformed signal for a representative simulation (\texttt{SXS:BBH:3982}) is shown in the top-left panel of Fig.~\ref{fig:fittingprocedure}. At low frequencies, the spectrum is dominated by the inspiral phase, where the amplitude follows the well-known post-Newtonian scaling $H_{\ell m}(\omega) \propto \omega^{-7/6}$ ~\cite{Finn:1992xs,Cutler:1994ys,Blanchet:2004ek}. This behavior sets a physical lower bound to the applicability of our ringdown model, which we therefore restrict to frequencies $\omega > \omega_x$, where $\omega_x$ is empirically determined from the data through the systematic fitting procedure described below. 

The spectrum $H_{\ell m}(\omega)$ decreases monotonically, approximately following a power-law behavior at intermediate frequencies
and an exponential decay at larger frequencies. Because of the rapid decay of the signal, $H_{\ell m}(\omega)$ eventually reaches the numerical noise level, beyond which the data are no longer physically meaningful.
To avoid contamination from numerical artifacts, the fits are performed only up to $\omega < \omega_{\rm cut}$, where $\omega_{\rm cut}$ is defined such that the numerical-noise contribution to
$H_{\ell m}(\omega)$ remains below $0.1\%$.  
A detailed discussion of how the cutoff frequency $\omega_{\rm cut}$ is determined
is given in Appendix~\ref{app:fft}; as we clarify there, the specific choice of $\omega_{\rm cut}$ does not affect the
estimation of the fitting parameters.

The fitting procedure proceeds as follows:

\begin{enumerate}
    \item We compute the Fourier transform of one of the numerical relativity simulations 
    from the SXS catalog, as described in Appendix~\ref{app:fft}, 
    and we discard it above the cutoff frequency $\omega_{\rm cut}$.
    The top-left panel of Fig.~\ref{fig:fittingprocedure} shows the resulting spectrum 
    for the simulation \texttt{SXS:BBH:3982}, focusing on the dominant $(\ell,m)=(2,2)$ mode.

    \item The remnant mass $M$ and dimensionless spin $\chi$ are available in the SXS metadata.
    Using these values, we compute the corresponding reflectivity $\mathcal{R}_{\ell m}(\bar{\omega},\chi)$
    and construct the reduced spectrum
    \[
    Y_{\ell m}(\omega) = \frac{H_{\ell m}(\omega)}{\sqrt{\mathcal{R}_{\ell m}(\bar{\omega},\chi)}}\,.
    \]

    \item We fit $Y_{\ell m}(\omega)$ with our model $M A_{\ell m}/\bar{\omega}^{p_{\ell m}}$.
Since the minimum frequency of validity of the model is not known a priori,
we consider a set of starting frequencies $\bar{\omega}_i$ such that 
$\bar{\omega}_i \in [0,\, \bar{\omega}_{\rm cut}]$.
The lowest frequency shown in Fig.~\ref{fig:fittingprocedure} is $\bar{\omega} = 0.2$,
which still lies within the inspiral region for the example simulation 
\texttt{SXS:BBH:3982}. 
To improve the numerical stability of the fit, we introduce a
simulation-dependent reference frequency $\bar{\omega}_*$ within the
fitting window, around which the power-law behavior is effectively anchored.
This is implemented by considering $(\bar{\omega}_*/\bar{\omega})^{p_{\ell m}}$
in the fit and subsequently reabsorbing the factor
$\bar{\omega}_*^{p_{\ell m}}$ into the amplitude.
This choice reduces correlations between the amplitude and the power-law
index and leads to more stable fits, without introducing any additional fitting
parameter.
In this work, we adopt the frequency $\bar{\omega}_*$ to be the knee frequency defined in
App.~\ref{app:hyperfitapplication}. Varying this choice within the same regime does not affect the inferred parameters.

    \item We then study the behavior of the fitted parameters $A_{\ell m}$ and $p_{\ell m}$
    as functions of the starting frequency $\bar{\omega}_i$.
    Within the validity region of the model, 
    the values of $A_{\ell m}$ and $p_{\ell m}$ obtained from different $\bar{\omega}_i$ are expected to be 
    consistent within their uncertainties.
    Outside this region, $A_{\ell m}$ and $p_{\ell m}$ vary significantly with $\bar{\omega}_i$, 
    as the model no longer reproduces the data accurately.
    As shown in the top-right panel of Fig.~\ref{fig:fittingprocedure} for $p_{\ell m}$
    and in the bottom-right panel for $A_{\ell m}$, 
    both parameters saturate to approximately constant values beyond a certain $\bar{\omega}_i$.
    To identify the frequency range where $A_{\ell m}$ and $p_{\ell m}$ are stable,
    we impose a maximum fractional variation of $0.1\%$ across the region.
    The frequency $\bar{\omega}_x$---the minimum frequency of validity of the model---is 
    then defined as the lowest $\bar{\omega}_i$ for which both $A_{\ell m}$ and $p_{\ell m}$ remain stable.

    \item Finally, we compare the fitted model with the data.
    In the top-left panel of Fig.~\ref{fig:fittingprocedure}, 
    the dashed pink line shows the best-fit model 
    for the simulation \texttt{SXS:BBH:3982}, 
    while the solid black line corresponds to the numerical spectrum.
    The bottom-left panel shows the mismatch, computed as in Eq.~\eqref{eq:mismatch},
    as a function of the starting frequency $\bar{\omega}_i$. 
    As expected, the mismatch reaches its minimum in the stability region, 
    with $\mathcal{M}\gtrsim10^{-5}$, 
    demonstrating an excellent agreement between the model and the numerical data.
\end{enumerate}

The fitting procedure described above can be applied to any simulation in the SXS catalog. 
Given the robustness of the proposed model, we now extend the analysis to a large set of simulations, 
in order to investigate how the fitted parameters $A_{\ell m}$ and $p_{\ell m}$ depend on the properties of the coalescing binary system.
As can be seen in Fig.~\ref{fig:fittingprocedure}, the greybody-based model shows good agreement with the spectrum even below the knee frequency (as defined in App.~\ref{app:hyperfitapplication}), often used as a phenomenological marker of the ringdown. This agreement should not be interpreted as evidence that the signal is already ringdown-dominated, but rather as an indication of the smooth onset of a Kerr-like perturbative response in the frequency domain.

\section{Modeling the parameter dependence of the fitted coefficients} \label{sec:fits}

We have shown how the quantities $A_{\ell m}$ and $p_{\ell m}$ can be extracted 
from the frequency-domain spectra of individual numerical relativity simulations. 
Our goal is now to model the dependence of these coefficients on the physical parameters of the binary system, following an approach similar to that adopted for the amplitudes and phases of QNMs in Ref.~\cite{Cheung:2023vki}.
For this part of the analysis we focus on nonprecessing, quasicircular BBHs. 
Within this class, we further select all SXS simulations with ID~$\geq 1000$ \footnote{Which typically correspond to more recent entries in the SXS catalog. This choice is intended as a pragmatic filter to avoid early legacy simulations, as more recent simulations benefit from later versions of the overall numerical infrastructure.}, 
and divide the resulting set into two disjoint subsets.
First, we randomly set aside about $\sim25\%$ of the simulations. This independent subset will later be used to assess the performance of our final model on data that never enter the fitting pipeline.
The remaining $\sim75\%$ of the simulations are used to model the dependence of the coefficients on the binary parameters.
This separation ensures independence between the construction of the model and the assessment of its validity.

The relevant physical parameters in our scenario are the component spins, $\chi_{1,z}$ and $\chi_{2,z}$ (for nonprecessing simulations the spins are either aligned or antialigned with the orbital angular momentum, 
so that $\vec{\chi}_1$ and $\vec{\chi}_2$ coincides with their components along the binary's angular momentum),
and the mass ratio $q=M_1/M_2$. It will be useful to employ related quantities that  often appear in post-Newtonian expansions, in particular 
\begin{equation}
   \chi_+ = \frac{q \chi_1+\chi_2}{1+q}, \qquad  \chi_- = \frac{q \chi_1-\chi_2}{1+q}\,, 
\end{equation}
represents the symmetric (antisymmetric) spin of the system, respectively, and 
\begin{equation}\label{eq:delta}
   \delta = \sqrt{1-4\eta}\quad\quad \text{with} \quad \eta={q \over (1+q)^2}\,.
\end{equation}
The mass asymmetry $\delta < 1$, and therefore $\delta$ provides a convenient variable for polynomial representations.

\begin{figure}[ht]
    \centering    
    \includegraphics[width=0.9\linewidth]{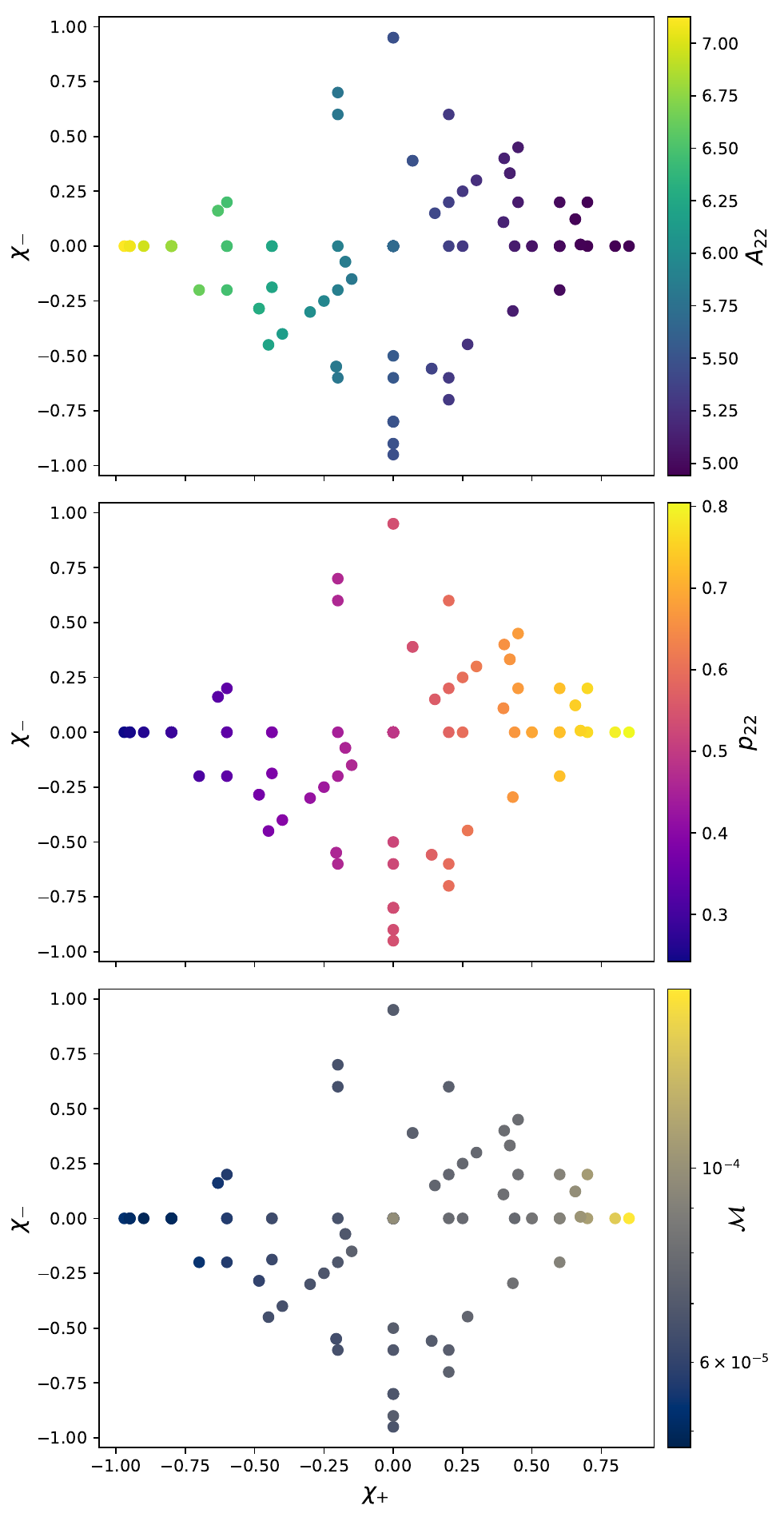} 
    \caption{
    Equal-mass slice for the $(\ell,m)=(2,2)$ coefficients.
    We select simulations with $q\in[0.999,\,1.001]$, corresponding to $\delta\simeq 0$,
    and show the variation of $A_{22}$ (top) and $p_{22}$ (middle) over the $(\chi_{+},\chi_{-})$ plane.
    Each point represents a numerical relativity simulation. The color scale encodes the value of the coefficients.
    The dependence on $(\chi_{+},\chi_{-})$ at fixed $\delta$ appears smooth. Moreover, the plot clearly shows that the dominant dependence is on $\chi_+$: variations in the values of $A_{22}$ and $p_{22}$ along $\chi_-$ (i.e., along vertical slices) are comparatively small.The bottom panel shows the corresponding mismatch between the numerical relativity simulations and our model, as defined in Eq.~\eqref{eq:mismatch}. All mismatches lie in the range $[4\times 10^{-5},\,2\times 10^{-4}]$. The largest mismatches are typically at higher $\chi_{+}$, where the impact of spherical-spheroidal harmonic mixing may become more significant.
}

    \label{fig:slice_q1}
\end{figure}

To gain preliminary insight into how the fitted coefficients depend on these quantities, we first examine a slice at nearly equal masses.
In particular, we select $100$ simulations with $q\in[1.000,\,1.001]$, which corresponds to
$\delta \simeq 0$, so that the dependence on the spins can be visualized cleanly in the
$(\chi_{+},\chi_{-})$ plane at essentially fixed $\delta$.
In Fig.~\ref{fig:slice_q1} we show the behavior of the $(\ell,m)=(2,2)$ coefficients, $A_{22}$ and $p_{22}$, across this slice.
Both quantities vary smoothly with the spins, providing a first indication that a low-order polynomial in $(\chi_+,\chi_-)$ should be adequate at $q\simeq 1$.

In general, our goal is to approximate the dependence of $A_{\ell m}$ and $p_{\ell m}$ by means of a polynomial model, in order to gain insight into how these quantities vary with the relevant physical parameters of the system. 
Of course, in full generality their functional dependence is not expected to be polynomial. 
However, since our analysis is restricted to a limited region of the parameter space $(\delta, \chi_+, \chi_-)$ corresponding to comparable-mass coalescences, a polynomial description can provide an accurate and tractable approximation.

For brevity, we denote by $X$ either of the two fitted quantities ($A_{\ell m}$ or $p_{\ell m}$) and, for any given choice of $(\ell,m)$, we model it as a multivariate polynomial of the form

\begin{equation}
X(\chi_1, \chi_2, \eta)
= \sum_{0 \le i,j,k \le \mathcal{N}} c_{ijk}\,
  \chi_1^{\,i}\,\chi_2^{\,j}\,\delta^{\,k},
\label{eq:poly_model}
\end{equation}
where $\mathcal{N}$ is the maximum degree of the polynomial.

Determining an appropriate value of $\mathcal{N}$ is nontrivial:
higher-order polynomials can overfit the data, while lower-order polynomials may fail 
to capture the relevant trends.
To better identify the optimal order, we employ a standard \textit{cross-validation} procedure~\cite{Hastie:2010}.

Specifically, we adopt a $k$-fold cross-validation scheme, in which the full set of simulations is partitioned into $k$ disjoint subsets. In this work we adopt $k=10$. At each iteration, one of the subsets is used as the \textit{validation set}, while the remaining $k-1$ subsets constitute the \textit{training set}.

The polynomial model is fitted on the training set, while the validation set is used to assess the quality of the fit, as discussed below. 

For each polynomial degree $\mathcal{N}$, we fit Eq.~\eqref{eq:poly_model} using the training set and compute the cross-validation mean squared error (CV~MSE) 
for both the training and validation data.
The CV~MSE is defined as
\begin{equation}
{\rm CV\,MSE}(\mathcal{N}) = 
\frac{1}{N_{\rm val}}
\sum_{i \in {\rm val}}
\big[ X_i^{\rm (data)} - X^{\rm (fit)}(\chi_{1,i}, \chi_{2,i}, \eta_i) \big]^2,
\label{eq:CVMSE}
\end{equation}
where $N_{\rm val}$ is the number of validation samples,
$X_i^{\rm (data)}$ is the measured value of $A_{\ell m}$ or $p_{\ell m}$ from the $i$-th simulation,
and $X^{\rm (fit)}$ is the corresponding prediction of the polynomial model.
The above procedure is repeated in a $k$-fold manner, such that each subset is used once as the validation set. The reported results are then obtained by averaging over the $k$ validation folds.

\begin{figure}[h]  \centering\includegraphics[width=\linewidth]{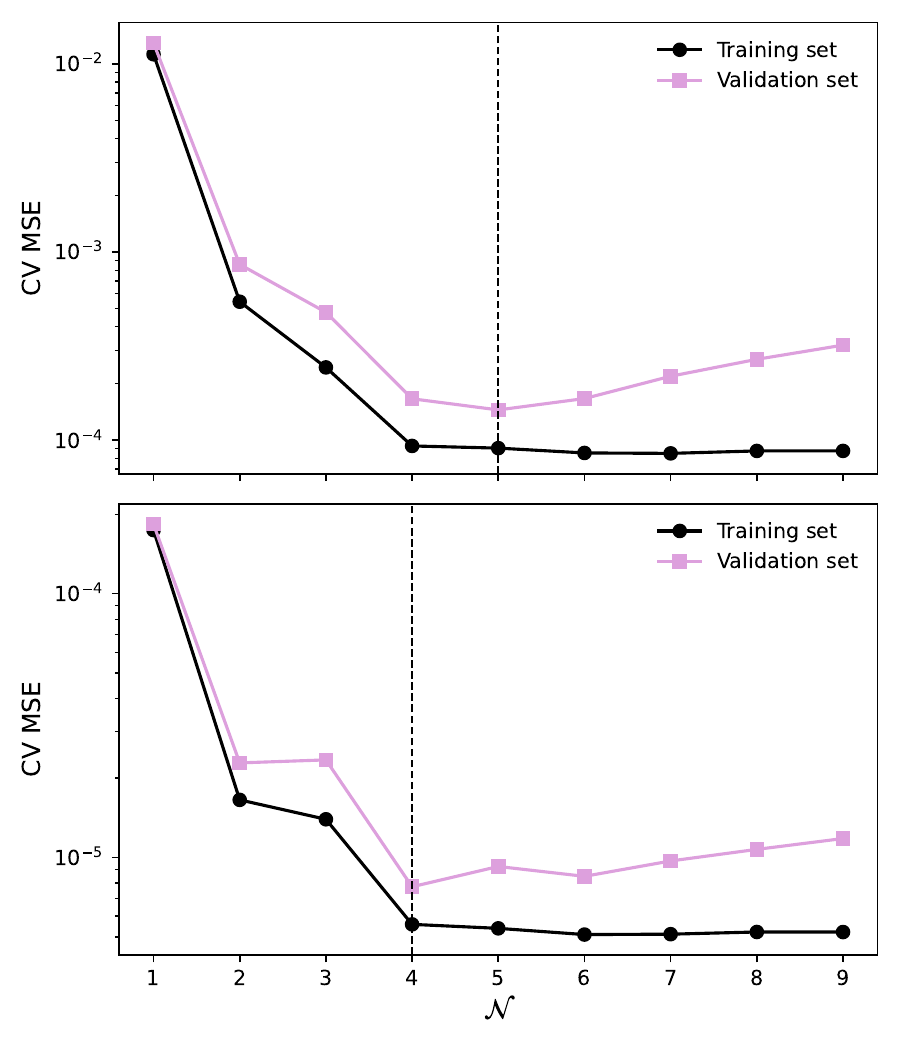}
    \caption{Cross-validation of the polynomial model for the fitted coefficients $A_{22}$ and $p_{22}$ for the $\delta=0$ slice of the $(2,2)$ mode, with $\delta$ defined in Eq.~\eqref{eq:delta}. 
The upper (lower) panel shows the cross-validation mean squared error (CV~MSE), defined in Eq.~\eqref{eq:CVMSE}, as a function of the polynomial degree $\mathcal{N}$ for $A_{22}$ ($p_{22}$). 
Results are displayed for both the training data (solid black) and the validation data (dashed pink), where the latter corresponds to the CV~MSE averaged over the $k$ validation folds}.
The training error decreases monotonically with $\mathcal{N}$, whereas the validation error reaches a minimum and then increases due to overfitting. 
The optimal degree $\mathcal{N}_{\rm opt}$, indicated by the vertical dashed line, is chosen as the value that minimizes the validation CV~MSE, thus providing the best compromise between accuracy and generality.
    \label{fig:parameterdependence}
\end{figure}

In general, the CV~MSE for the training set decreases monotonically with the polynomial degree $\mathcal{N}$, since higher values of $\mathcal{N}$ mean that more coefficients are available to fit the data. This, however, can lead to overfitting. The validation error, being fully independent of the training set, typically exhibits a minimum that signals the onset of overfitting.
We therefore identify the optimal degree $\mathcal{N}$ as the one that minimizes the validation CV~MSE, ensuring an accurate model without overfitting.

To illustrate the method more clearly, we apply the
procedure to the $(2,2)$ mode on the slice $\delta = 0$ shown in Fig.~\ref{fig:slice_q1}, where the
dependence reduces to two dimensions and is therefore easier to visualize.  
The same procedure will then be applied in full generality to the complete
three-dimensional parameter space $(\delta, \chi_+, \chi_-)$.

We consider the training set and validation set defined above. We then compute the CV~MSE for polynomials of increasing total
degree $\mathcal{N}$, and identify the optimal degree as the value that
minimizes the validation error. The results, shown in
Fig.~\ref{fig:parameterdependence}, exhibit the expected behavior: the CV~MSE
for the training set (black dots) decreases monotonically with $\mathcal{N}$,
while the validation error (pink squares) develops a clear minimum at optimal degrees of $\mathcal{N}=4$ for $p_{22}$ and
$\mathcal{N}=5$ for $A_{22}$.

However, a polynomial of total degree $\mathcal{N}$ in $d$ variables contains
$\binom{\mathcal{N}+d}{\mathcal{N}}$ independent coefficients. Thus, fixing only
the degree $\mathcal{N}$ would in general allow for a large number of
terms. To control this, we apply a second selection step: we rerun the CV~MSE
algorithm over all possible subsets of monomials of degree up to
$\mathcal{N}$, and choose the configuration that minimizes the validation CV
MSE. Applying this procedure to the $\delta = 0$ slice for the $(2,2)$ mode, we
obtain
\begin{equation}
\begin{aligned}
&A_{22}(\chi_+,\chi_-,\delta=0) \simeq\;
  5.620
- 1.346\,\chi_+
+ 0.209\,\chi_+^{2} \\
&- 0.261\,\chi_-^{2}
+ 0.289\,\chi_+^{3}
+ 0.282\,\chi_+^{4} 
+ 0.291\,\chi_+ \chi_-^{2} \\
&+ 0.120\,\chi_-^{4}
+ 0.779\,\chi_+^{2} \chi_-^{2}
- 0.077\,\chi_+\chi_-^{4}\,,
\end{aligned}
\end{equation}
and
\begin{equation}
\begin{aligned}
&p_{22}(\chi_+,\chi_-,\delta=0) \simeq\;
  0.509
+ 0.339\,\chi_+
+ 0.088\,\chi_+^{2}\\
& 
+ 0.057\,\chi_-^{2}
- 0.037\,\chi_+^{3}
- 0.059\,\chi_+^{4} - 0.025\,\chi_-^{4}\\
& 
- 0.144\,\chi_+^{2}\chi_-^{2}
- 0.027\,\chi_+\chi_-^{2}\,.
\end{aligned}
\end{equation}

\begin{figure}[t]
    \centering    
    \includegraphics[width=0.48\textwidth]{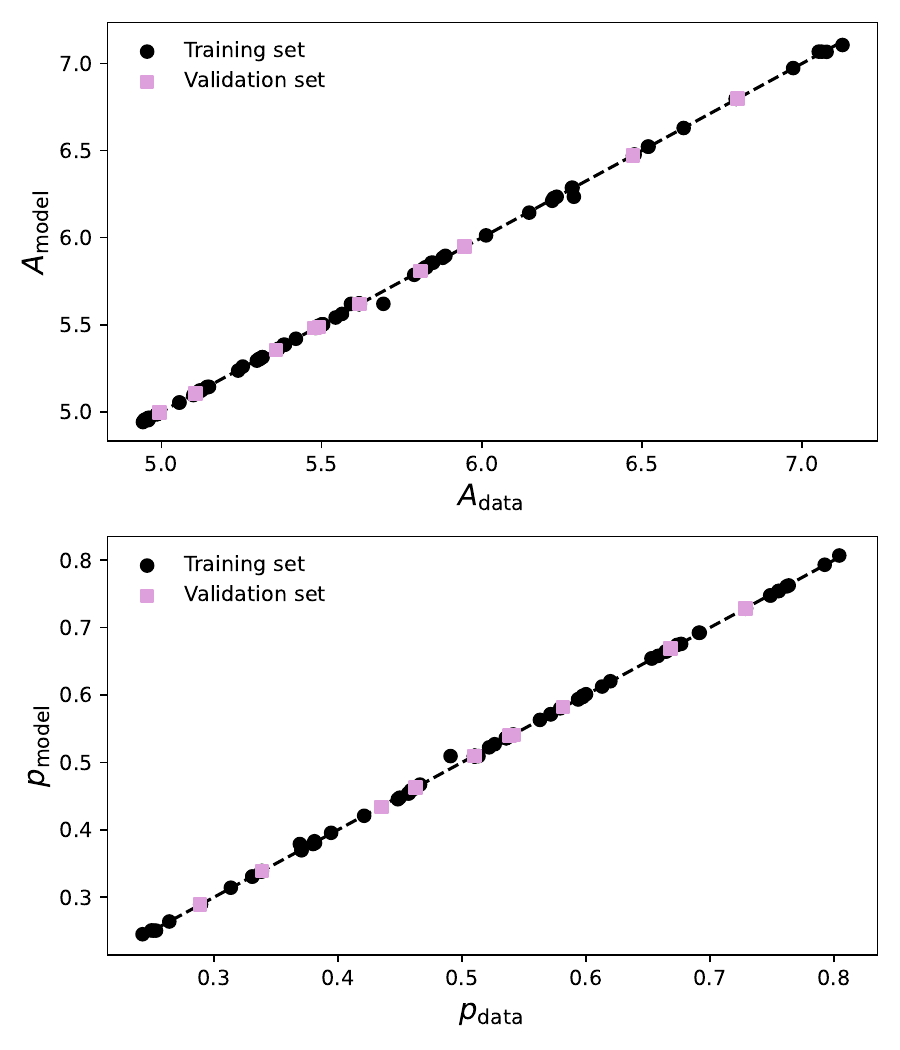}
    \caption{
    Comparison between the values predicted by the polynomial model of Eq.~\eqref{eq:poly_model}, $X_{\mathrm{fit}}$, and the corresponding values directly extracted from the simulations, $X_{\mathrm{data}}$, through the fitting procedure described in Sec.~\ref{subsec:fittingprocedure}. The top panel refers to $A_{22}$, and the bottom panel to $p_{22}$. Black (pink) points correspond to the training (validation) data for a representative iteration of the $k$-fold cross-validation procedure}. The dashed line shows $X_{\mathrm{fit}} = X_{\mathrm{data}}$, along which both datasets lie, indicating a good fit with no evidence of overfitting.
    \label{fig:fit_quality}
\end{figure}
The behavior observed in Fig.~\ref{fig:slice_q1} suggests that both $A_{22}$ and $p_{22}$ should depend primarily on $\chi_{+}$, since no significant variation is visible when moving along $\chi_{-}$. This expectation is fully confirmed by the fit result. At linear order, both quantities depend exclusively on $\chi_{+}$, with no contribution from $\chi_{-}$. The latter enters only at higher orders, and through even powers only. This ensures that $A_{22}$ and $p_{22}$ remain invariant under the exchange $\chi_{1,z} \leftrightarrow \chi_{2,z}$, as required by the symmetry of an equal-mass binary under the permutation of the two bodies. 
To assess the accuracy of the polynomial model, in Fig.~\ref{fig:fit_quality} we compare the values of $X_{22}$ (i.e., $A_{22}$ and $p_{22}$) extracted directly from the simulations, $X_{\mathrm{data}}$, with the corresponding predictions of the best-fit model of Eq.~\eqref{eq:poly_model}, $X_{\mathrm{fit}}$.  
The points lie very close to the diagonal $X_{\mathrm{fit}} = X_{\mathrm{data}}$, indicating good agreement between the model and the data. Both the training set and the validation set exhibit agreement, confirming that the model generalizes well and does not suffer from overfitting.

The same fitting strategy can be extended to capture the full three-parameter dependence.  
In Appendix~\ref{app:fitresults} we list the resulting fits for several $(\ell,m)$ modes and show the analog of Fig.~\ref{fig:fit_quality}, which we use to assess their quality. We also corroborate these fits using the results of an independent Random Forest Regressor (RFR).
As expected, moving to a three-parameter model increases the complexity of the problem, and the fits tend to become less accurate when evaluated across the entire parameter space.
This deterioration is particularly visible in two cases. First, modes with intrinsically small amplitudes--such as the $(3,3)$ mode near the $\delta=0$ slice--are more challenging to model reliably, as the numerical extraction of their Fourier transforms becomes less precise. Second, the poorest fits are observed for the $(4,4)$ mode, whose higher characteristic frequencies make it more sensitive to spin-induced spheroidal-spherical harmonic coupling. A proper treatment of this effect is an interesting topic for follow-up work.

\begin{figure}[h]
    \centering    \includegraphics[width=\linewidth]{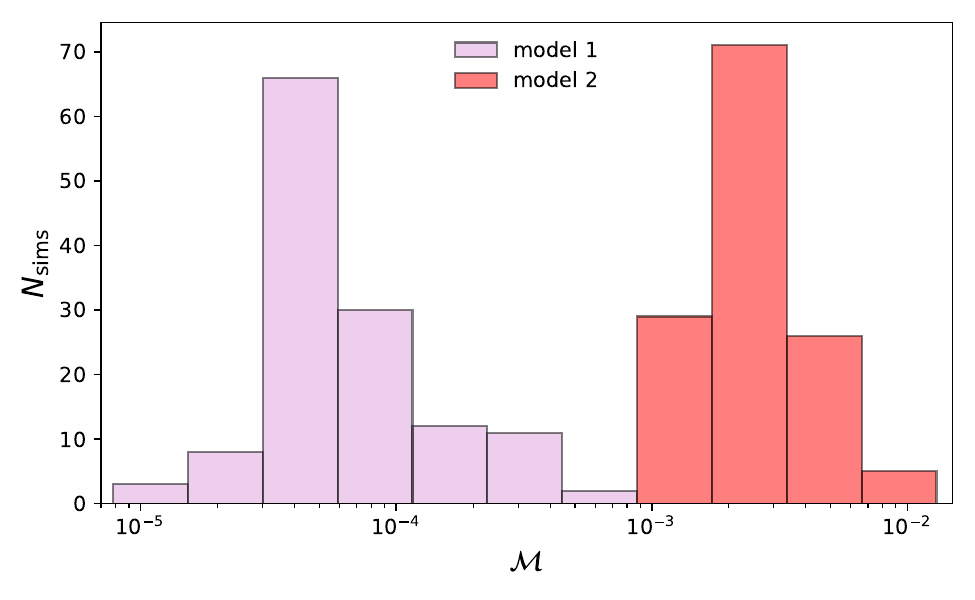}
    \caption{%
    Distribution of mismatches between the numerical SXS $(2,2)$ modes and the two models discussed in  Sec.~\ref{sec:modelcomparison}. Model $1$ refers to $H^{(1)}=MA_{22}\,\sqrt{\mathcal{R}_{22}}$, and model $2$ to $H^{(2)}=MA_{22}\,\sqrt{\mathcal{R}_{22}}/\bar{\omega}^{p_{22}}$.  
Both models are evaluated on a separate set of simulations (ID $>1000$, $\eta\in[0.15,0.25]$, circular, nonprecessing), which was not used at any stage of the fitting procedure.  
This dataset, completely independent from both the training and the 
validation sets, allows us to test the true predictive performance of the models.
For each simulation, the mismatch is computed using Eq.~\eqref{eq:mismatch}.
The model $H^{(2)}$ of Eq.~\eqref{eq:model}, with parameters
$(A_{22},p_{22})$ obtained from the hyper-fit formulas of
Appendix~\ref{app:fitresults}, yields mismatches that remain below $10^{-3}$
for all cases, and typically at the level of $\mathcal{O}(10^{-5}-10^{-4})$.  
In contrast, the simpler model $H^{(1)}$ produces a
distribution shifted toward larger values, with mismatches generally one to
two orders of magnitude higher.  
This confirms the superior accuracy of the $H^{(2)}$ model in reproducing the
numerical SXS $(2,2)$ data across the parameter space.
} \label{fig:mismatch_histogram}
\end{figure}

It is instructive to comment more specifically on the performance of the $(2,2)$ results, 
which represent the most reliable part of the analysis, as this mode dominates the 
gravitational-wave signal. To this purpose, we test the model on a set of simulations entirely independent of both the training and validation datasets, i.e., the subset that was set aside before the fitting procedure.  
For each simulation, we compute the mismatch between the numerical $(2,2)$ mode and the model of Eq.~\eqref{eq:model}, using the values of $A_{22}$ and $p_{22}$ predicted by the hyper-fit formulas of Appendix~\ref{app:fitresults}.  
We then compare these results with those obtained using the alternative model $H^{(1)} = M A\,\sqrt{\mathcal{R}_{\ell m}}$ discussed in Sec.~\ref{sec:modelcomparison}, for which we repeat the full fitting procedure described in the paper.  
The resulting mismatch distributions are shown in Fig.~\ref{fig:mismatch_histogram}, 
with the model $H^{(2)}$ displayed in pink and the model $H^{(1)}$ in red. In App.~\ref{app:AdvLIGO} we recompute the mismatches using the Advanced LIGO sensitivity curve.

As shown in Fig.~\ref{fig:mismatch_histogram}, for the model $H^{(2)}$ analyzed throughout the paper the mismatch always remains below $10^{-3}$.  
The upper bound is reached only when the binary parameters lie near the edges of the explored domain, in particular for high values of $\chi_\pm$, as discussed in Appendix~\ref{app:fitresults}. For most simulations the mismatch is of order $\mathcal{O}(10^{-5}-10^{-4})$.  
By contrast, the mismatch distribution obtained using the model $H^{(1)}$ is systematically shifted towards larger values--typically by one to two orders of magnitude.  
This confirms the intuition of Sec.~\ref{sec:modelcomparison}: the second model provides a much more accurate description of the numerical data.

In order to further validate the robustness of the fitted models, we also perform an independent regression analysis using an RFR, as discussed in Appendix~\ref{app:fitresults}. 
In general, the RFR results suggest that polynomials of the form~\eqref{eq:poly_model} provide good fits to the data, as can be seen in Fig.~\ref{fig:rfr_vs_poly} of the same appendix.

The RFR also provides some insight with respect to optimal parameterizations of $A_{\ell m}$ and $p_{\ell m}$ in terms of physical parameters. We use the RFR described in Appendix~\ref{app:fitresults} to fit these parameters as functions of $(\delta,\chi_+,\chi_-)$, $(\delta,\chi_{1,z},\chi_{2,z})$, and $(M,\chi)$ (i.e., the remnant parameters). We summarize some results in Table~\ref{tbl:feature_importance}, where we list the mean absolute error and R-squared score as metrics for assessing the goodness of fit with the different parameterizations. These metrics offer a concise evaluation of the fitting quality: an accurate model exhibits a mean absolute error approaching zero and an R-squared score close to one. Here, we see that the two parameterizations involving progenitor parameters are quite competitive, while the remnant parameters give worse fits of $A_{\ell m}$ and $p_{\ell m}$, sometimes dramatically so. Note that there are considerably fewer simulations with an appreciable amplitude to be fitted for the $(2,1)$ and $(4,4)$ modes (training sets of 96 and 60 simulations, respectively), compared to the $(2,2)$ and $(3,3)$ modes (278 and 127). This reduced sampling limits the constraining power of the polynomial regressor and results in a visibly larger scatter in the fitted values for these subdominant modes, as shown in Fig.~\ref{fig:rfr_vs_poly}. 
This effect should therefore be taken into account when comparing the goodness of fit between different modes, as a larger number of simulations generally yields a more accurate and stable fit.

\begin{table}[t]
\caption{Relative performance of three different parameterizations of $A_{\ell m}$ and $p_{\ell m}$ as determined by the RFR. The SXS simulations used all have $\eta > 0.15$; for the $(3,3)$ and $(2,1)$ modes, we further restrict to $\eta < 0.25$, given the smallness of the frequency-domain amplitudes at $\eta=0.25$. }
\tabcolsep=0.31cm
\begin{tabular}{c c c c}
 \hline
 \hline
\multicolumn{4}{c}{Mean Absolute Error} \\
 \hline
  & $\delta,\chi_+,\chi_-$ &$\delta,\chi_{1,z},\chi_{2,z}$ &$M,\chi$\\
 \hline
 $A_{22}$& 0.051 & 0.047 & 0.069 \\
 $p_{22}$& 0.010 & 0.015 & 0.008 \\
 $A_{33}$& 0.010 & 0.013 & 0.035 \\
 $p_{33}$& 0.036 & 0.033 & 0.054 \\
 $A_{21}$& 0.129 & 0.128 & 0.254 \\
 $p_{21}$& 0.018 & 0.015 & 0.025 \\
 $A_{44}$& 0.035 & 0.042 & 0.052 \\
 $p_{44}$& 0.721 & 0.606 & 0.964 \\
 \hline
 \hline
 \multicolumn{4}{c}{R-squared Score} \\
 \hline
  & $\delta,\chi_+,\chi_-$ &$\delta,\chi_{1,z},\chi_{2,z}$ &$M,\chi$\\
 \hline
 $A_{22}$& 0.975  & 0.984  & 0.955 \\
 $p_{22}$& 0.982  & 0.971  & 0.989 \\
 $A_{33}$& 0.997  & 0.996  & 0.951 \\
 $p_{33}$& 0.810  & 0.877  & 0.500 \\
 $A_{21}$& 0.934  & 0.939  & 0.721 \\
 $p_{21}$& 0.944  & 0.961  & 0.872 \\
 $A_{44}$& 0.700  & 0.576  & 0.079 \\
 $p_{44}$& -0.478 & 0.056  & -1.831\\
 \hline
\end{tabular}
\label{tbl:feature_importance}
\end{table}

\begin{figure}[h]
    \centering
    \includegraphics[width=0.48\textwidth]{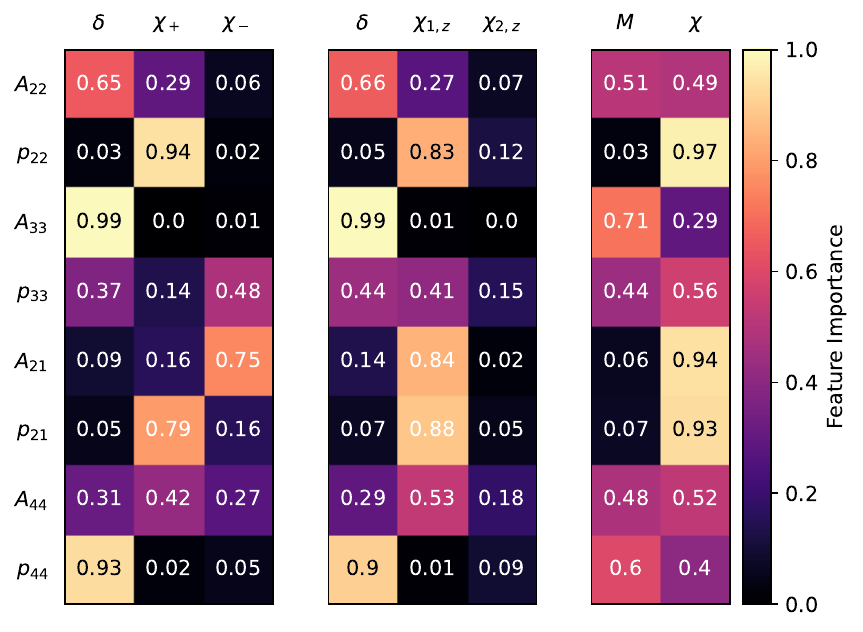}
    \caption{Relative ``importance'' of different BBH parameters, as determined by the random forest regressor.}
    \label{fig:feature_importance}
\end{figure}

In addition to helping us determine which is the optimal parameterization between the three, the RFR also provides some indication of which physical parameters most influence the value of $A_{\ell m}$ and $p_{\ell m}$ within a single parameterization. This intuition is provided through the \texttt{feature\_importances\_} attribute of the regressor described in Appendix~\ref{app:fitresults}. This attribute assigns a relative ``importance'' (quantified as a weight between 0 and 1) to each fitting parameter; these weights add to one for each parameterization. In Fig.~\ref{fig:feature_importance}, we show the feature importance between individual parameters for each attempted parameterization. For the (2,~2) and (3,~3) modes, the mass parameters carry more weight in determining $A_{\ell m}$, while the spin parameters tend to carry more weight in determining $p_{\ell m}$. For the (2,~1) mode, there is little dependence on the mass for either $A_{21}$ or $p_{21}$. We note that spins enter at lower order in the post-Newtonian terms for the (2,~1) mode than for either the (2,~2) or (3,~3). 

\section{Discussion} \label{sec:discussion}
In this work we have developed and validated a frequency-domain model for the ringdown of comparable-mass BBH mergers based on BH greybody factors (and specifically on the remnant reflectivity). Our main result is that the model of Eq.~\eqref{eq:model} provides an accurate and robust description of the amplitudes of the numerical spectra extracted from the SXS catalog. When evaluated on the optimal frequency range identified through our stability analysis, the model achieves mismatches as low as $\mathcal{O}(10^{-5})$, improving previous formulations~\cite{Okabayashi:2024qbz} by more than one order of magnitude.

As a useful rule of thumb, for a signal with a given SNR, two waveforms are deemed indistinguishable if their mismatch~\cite{Flanagan:1997kp,Lindblom:2008cm}
\begin{equation}
  \mathcal{M} \lesssim \frac{1}{2\,{\rm SNR}^2}\,.
\end{equation}
Since this condition is sufficient but not necessary, it typically provides a conservative, and often overly restrictive, criterion (see, e.g.~\cite{Chandramouli:2024vhw}). Using the above estimate, a mismatch ${\cal M}=\mathcal{O}(10^{-5})$ would guarantee no modeling systematics for post-merger SNR as high as $\approx 200$, well beyond current capabilities.

We have shown that the parameters $(A_{\ell m},p_{\ell m})$ vary smoothly across 
the physical parameter space $(\delta,\chi_+,\chi_-)$ of comparable-mass 
mergers, and that their dependence can be efficiently captured by low-degree 
multivariate polynomials selected through a dedicated cross-validation 
procedure. The comparison with a fully nonparametric 
RFR further confirms the robustness of our polynomial 
representation and highlights the physical relevance of the chosen variables, 
with $\chi_+$ and $\delta$ emerging as the dominant predictors for most modes.

Overall, our analysis demonstrates that BH reflectivity provides a powerful and physically motivated framework for modeling the ringdown of comparable-mass mergers directly in the frequency domain. These results open the possibility of extracting the remnant properties $(M,\chi)$ from the ringdown independently of a traditional QNM analysis, thereby enabling new and complementary strong-gravity tests with future gravitational-wave observations. 

We emphasize that the reflectivity-based framework can be employed in two conceptually distinct ways.
In a first, fully agnostic approach, the amplitude and the power-law index $(A_{\ell m},p_{\ell m})$ are treated as free fitting parameters. In this case, the remnant mass and spin $(M,\chi)$ can be inferred directly from the ringdown signal, without imposing any model for the progenitor properties (see Fig.~\ref{fig:modelcomparison}).
Alternatively, one may adopt a model for $(A_{\ell m},p_{\ell m})$ that encodes their dependence on the progenitor parameters.
Within this approach, the same framework allows one to propagate information from the inspiral-merger phase into the ringdown, enabling the extraction of progenitor properties through their imprint on the Fourier-domain amplitude of the remnant waveform.

Finally, the reflectivity-based framework naturally lends itself to a number of extensions, including the modeling of the effect of orbital configurations preceding ringdown, such as eccentricity and spin precession~\cite{Anselmo:2025ehx}, and the inclusion of spin-induced spherical-spheroidal mode coupling~\cite{Berti:2014fga}. 
A natural extension is the application of the model to real data; this would require modeling the signal phase in Fourier space in addition to its amplitude.
These developments would allow for a more faithful description, and they will be the subject of future work.

\begin{acknowledgments}
We thank Mark H.-Y. Cheung for insightful conversations on the polynomial fits, and Jay Wadekar for inspiration and guidance in using the Random Forest Regressor for the fits. 
R.F.R. thanks A.~Scarpa and F.~Crescimbeni for useful suggestions and discussions.
R.F.R. and P.P. are supported by the MUR FIS2 Advanced Grant ET-NOW (CUP:~B53C25001080001) and by the INFN TEONGRAV initiative. S.Y. and E.B. are supported by NSF Grants No.~AST-2307146, No.~PHY-2513337, No.~PHY-090003, and No.~PHY-20043, by NASA Grant No.~21-ATP21-0010, by John Templeton Foundation Grant No.~62840, by the Simons Foundation [MPS-SIP-00001698, EB], by the Simons Foundation International [SFI-MPS-BH-00012593-02], and by Italian Ministry of Foreign Affairs and International Cooperation Grant No.~PGR01167. S.Y. is supported by the NSF Graduate Research Fellowship Program under Grant No. DGE2139757.
Some numerical computations were performed at the Vera cluster, supported by MUR and Sapienza University of Rome.

\end{acknowledgments}

\appendix

\section{Numerical Fourier transform}\label{app:fft}
In performing the Fourier transform of SXS simulations, we use the standard preprocessing steps available through the \texttt{sxs} python package. The necessity and implementation of these steps are explained in Ref.~\cite{Chen:2024ieh}. They include padding, tapering, line subtracting, and transitioning to a constant in such a way that the desired features in the signal (e.g., gravitational-wave memory) are minimally affected. While the results in the main text are obtained using default settings in this preprocessing step, we have also checked that alternative choices for padding, tapering, etc. generally do not make a significant difference in the final fits for $A_{\ell m}$ and $p_{\ell m}$. 
\begin{figure}[h]
    \centering
    \includegraphics[width=0.4\textwidth]{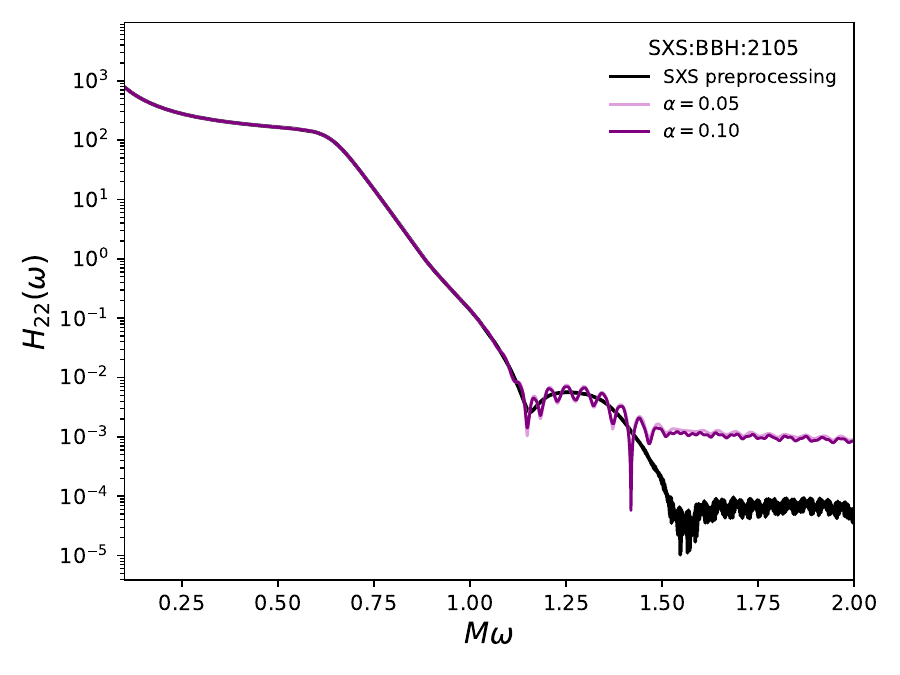}
    \caption{
        Comparison between the FFT of the preprocessed SXS waveform (black curve) and the FFTs computed from the $(2,2)$ mode of the simulation \texttt{SXS:BBH:2105} after removing the junk radiation and applying a Tukey window with windowing parameters $\alpha=0.05$ (pink) and $\alpha=0.10$ (purple).  
        All spectra agree closely except at the highest frequencies, where numerical noise dominates. The shaded band marks the region where the noise floor reaches approximately $0.1\%$ of the signal amplitude, which determines the upper cutoff frequency used in the fitting procedure. }
    \label{fig:fftprepvstaper}
\end{figure}

In Fig.~\ref{fig:fftprepvstaper} we compare the Fourier transform obtained using the \texttt{sxs} preprocessing routine with the FFT computed directly from the $(2,2)$ mode after removing the early junk radiation, interpolating to a uniform time grid, performing zero-padding, and applying a Tukey window with two different tapering parameters $\alpha$. As shown in the figure, all three spectra agree extremely well over the entire frequency range relevant for the fit. Differences appear only at the highest frequencies, where the spectrum becomes dominated by numerical noise.  Extracting the parameters $A_{\ell m}$ and $p_{\ell m}$ from each of the three spectra yields consistent results within numerical accuracy, confirming that the fitting procedure is robust against the details of the FFT pipeline. In all cases, a noise floor is present at high frequencies. This noise floor sets the maximum frequency up to which the fit can be performed. In practice, we truncate the spectrum at a frequency where the noise floor reaches $0.1\%$ of the signal amplitude. We have verified that varying this threshold (in the range $0.1-5.0\%$) does not lead to any appreciable change in the best-fit values of $A_{\ell m}$ and $p_{\ell m}$.

\section{Reflectivity from the Teukolsky equation}
\label{app:greybody_teukolsky}
We summarize here the standard procedure to compute the reflectivity
$\mathcal{R}_{\ell m}(\bar{\omega},\chi)$ for gravitational perturbations of a Kerr
BH. In Boyer-Lindquist coordinates, the Kerr metric reads
\begin{align}
&ds^2 = -\left(1-\frac{2Mr}{\Sigma}\right)dt^2 
       + \frac{\Sigma}{\Delta}dr^2 
       - \frac{4Mar\sin^2\theta}{\Sigma}\,dt\,d\phi \notag\\
& + \Sigma\,d\theta^2
       + \left[\frac{(r^2+a^2)^2 - a^2\Delta\sin^2\theta}{\Sigma}\right]
         \sin^2\theta\,d\phi^2 ,
\end{align}
where $\Sigma=r^2+a^2\cos^2\theta$, $\Delta=r^2+a^2-2Mr$, and the horizon is
at $r_+=M+\sqrt{M^2-a^2}$.  Scalar, electromagnetic and gravitational
perturbations are described by the Teukolsky master equations
\cite{Teukolsky:1972my}.  After separation of
variables, the radial function ${}_sR_{\ell \bar{\omega}}(r)$ satisfies
\begin{align}
&\Delta^{-s}\frac{d}{dr}\!\left(\Delta^{s+1}\frac{d{}_s R}{dr}\right) 
+ \Bigg[\frac{K^2}{\Delta} +\quad\quad\quad\nonumber\\&\quad-\frac{2is(r-M)K}{\Delta} +4is\omega r - \lambda\Bigg]{}_s R=0,
\end{align}
where $K=(r^2+a^2)\omega-am$, $s$ is the spin weight, and $\lambda$ is the
angular separation constant determined by the spin-weighted spheroidal
harmonics. Introducing the tortoise coordinate $dr_*/dr=(r^2+a^2)/\Delta$,
the horizon corresponds to $r_*\to-\infty$ and spatial infinity to
$r_*\to+\infty$.  For fixed $(\ell,m,\omega)$, the physical solution is
purely ingoing at the horizon,
\begin{equation}
{}_sR(r) \xrightarrow[r\to r_+]{} \Delta^{-s} e^{-ik r_*},
\qquad k=\omega-m\Omega_{\rm H} ,
\end{equation}
where $\Omega_H$ is the angular velocity of the horizon.  At large radii,
the solution behaves as
\begin{equation}
{}_sR(r)\xrightarrow[r\to\infty]{}
Z^{\rm out}_{\ell \bar{\omega}}\, r^{-1-2s}e^{+i\omega r_*}
+ Z^{\rm in}_{\ell \bar{\omega}}\, r^{-1}e^{-i\omega r_*},
\end{equation}
which for gravitational perturbations ($s=-2$) reduces to
\(
{}_sR\sim Z^{\rm out} r^{3}e^{+i\omega r_*}+Z^{\rm in} r^{-1}e^{-i\omega r_*}.
\)
The amplitudes $Z^{\rm in}$ and $Z^{\rm out}$ are obtained by matching the
numerical solution of the Teukolsky equation with its asymptotic form~\cite{Pani:2013pma}.  
Because the Teukolsky function does not carry unit flux, one must convert
$(Z^{\rm in},Z^{\rm out})$ into physical reflection and transmission
coefficients.  The required normalization factors were derived by Teukolsky
and Press~\cite{Press:1972zz,Press:1973zz} and independently by Starobinsky and
Churilov~\cite{Starobinskii:1973vzb,Starobinskil:1974nkd}.  For spin $s=-2$
gravitational perturbations, the reflectivity takes the form
\begin{equation}
\mathcal{R}_{\ell m}( \bar{\omega},\chi)
  = \frac{C}{256\,\omega^8}
    \left|\frac{Z^{\rm out}_{\ell m\omega}}{Z^{\rm in}_{\ell m\omega}}\right|^2 ,
\end{equation}
where the normalization constants $B$ and $C$ are
\begin{align}
B &= \left(\lambda + s(s+1)\right)^{2}
     + 4 m a \omega
     - 4 a^{2}\omega^{2},
\\[4pt]
C &= B\Big[\big(\lambda + s(s+1) -2\big)^{2}
              + 36 a\omega m
              - 36 a^{2}\omega^{2}\Big]
\nonumber\\
&\quad + \left[2\big(\lambda + s(s+1)\big) -1\right]
       \left(96 a^{2}\omega^{2} - 48 a\omega m\right)
\nonumber\\
&\quad + 144\,\omega^{2}(M^{2}-a^{2}).
\end{align}
These coefficients encode the flux-normalization appropriate for the
Teukolsky radial function, ensuring that $\mathcal{R}_{\ell m}( \bar{\omega},\chi)=1$
corresponds to perfect reflection and $\mathcal{R}_{\ell m}( \bar{\omega},\chi)=0$ to
perfect absorption by the BH.

The reflectivity, and therefore the greybody factors through Eq.~\eqref{eq:gfsrefl}, is naturally defined on a spheroidal harmonic angular basis. By contrast, SXS waveforms are decomposed in spherical harmonics. Switching between these two bases induces a mixing among modes, with coefficients that scale with $a\omega$~\cite{Berti:2005gp,Berti:2014fga}. In this analysis we neglect such mixing, which is expected to be small in the frequency range of interest. A more accurate treatment will nevertheless require incorporating this coupling, especially to improve the agreement at high spins and for higher-$\ell$ multipoles.

\section{Fit results and validation}\label{app:fitresults}
In this section we present the results of the polynomial fits for the mode amplitudes $A_{\ell m}$ and spectral slopes $p_{\ell m}$ across the parameter space $(\delta, \chi_{+}, \chi_{-})$.  
For each mode, we report the optimal polynomial structure selected by our cross-validation procedure and compare the fitted values with those extracted directly from the numerical simulations.  
This allows us to assess both the accuracy and the generality of the model. 

For modes with odd $m$ we adopt a slightly different ansatz for the amplitudes 
with respect to Eq.~\eqref{eq:poly_model}, since these terms are expected to 
vanish in the $\delta=0$ limit. Motivated by post-Newtonian inspired models, we follow the 
same structure used for QNM amplitudes in Ref.~\cite{Cheung:2023vki}, namely
\begin{equation}
    A_{\ell m}(\chi_{+},\chi_{-},\delta)
    = b\,|\chi_{-}|
    + \delta \sum_{0\leq i,j,k\leq \mathcal{N}}
      c_{ijk}\,\chi_{+}^{i}\chi_{-}^{j}\delta^{k}\,.
\end{equation}

To complement this analysis and further validate the robustness of our results,
we also perform an independent, nonparametric regression using a Random Forest
Regressor (RFR).  
The RFR predictions, presented mode by mode, provide a fully
model-independent benchmark that confirms the reliability of the polynomial
fits over the explored region of the parameter space.

For the $(2,2)$ mode we obtain the following best-fit expression:

\vspace{-1 cm}
\begin{widetext}
\begin{equation}
\begin{aligned}
&A_{22}(\delta,\chi_{+},\chi_{-}) \simeq\;
5.620
- 1.346\,\chi_{+}
+ 0.209\,\chi_{+}^{2}
- 0.261\,\chi_{-}^{2}
+ 0.289\,\chi_{+}^{3}
+ 0.282\,\chi_{+}^{4}
+ 0.291\,\chi_{+}\chi_{-}^{2}
+ 0.120\,\chi_{-}^{4} \\ &
+ 0.779\,\chi_{+}^{2}\chi_{-}^{2}
- 2.932\,\delta^{2}
- 0.776\,\chi_{+}\delta^{2}
- 1.864\,\chi_{+}\chi_{-}\delta
- 2.020\,\chi_{-}\delta
- 5.220\,\delta^{4} 
- 2.347\,\chi_{+}^{2}\delta
+ 2.619\,\chi_{+}^{2}\delta^{2}\\ &
- 1.993\,\chi_{+}^{2}\chi_{-}\delta
+ 4.212\,\chi_{-}\delta^{3}
+ 1.515\,\chi_{+}\delta 
+ 2.025\,\chi_{+}\delta^{3}
+ 1.577\,\chi_{+}\chi_{-}\delta^{2}
- 0.198\,\chi_{-}^{2}\delta
- 0.077\,\chi_{+}\chi_{-}^{4}\,.
\end{aligned}
\end{equation}
\end{widetext}
\begin{widetext}
\begin{equation}
\begin{aligned}
&p_{22}(\chi_+,\chi_-,\delta) \simeq\;
  0.509
+ 0.339\,\chi_+
+ 0.091\,\chi_+^{2}
+ 0.060\,\chi_-^{2}
- 0.039\,\chi_+^{3}
- 0.064\,\chi_+^{4}
- 0.030\,\chi_-^{4} - 0.156\,\chi_+^{2}\chi_-^{2}\\&
- 0.030\,\chi_+\chi_-^{2}
- 0.083\,\delta
+ 0.497\,\chi_+^{2}\delta
+ 0.051\,\chi_-^{3}\delta + 0.202\,\chi_-^{2}\delta
- 0.260\,\chi_+ \delta
+ 0.776\,\chi_+^{2}\chi_- \delta
+ 0.265\,\chi_- \delta \\
&- 1.266\,\chi_- \delta^{3}
- 0.725\,\chi_+^{2}\delta^{2}
+ 0.471\,\chi_+ \chi_- \delta
+ 2.099\,\delta^{4}- 1.372\,\delta^{3}
- 0.352\,\chi_-^{2}\delta^{2}
+ 0.417\,\chi_- \delta^{2}
- 0.134\,\chi_+ \delta^{3} \,.
\end{aligned}
\end{equation}
\end{widetext}
\newpage
For the $(3,3)$ mode we obtain 
\begin{widetext}
\begin{equation}
\begin{aligned}
&A_{33}(\delta,\chi_{+},\chi_{-}) 
\simeq 
0.0158\,|\chi_{-}|+\delta \left(
+ 3.770
- 3.943\,\delta^{2}
- 0.198\,\chi_{+}
+ 0.956\,\chi_{+}\delta^{3}
+ 1.347\,\chi_{-}^{2}\delta^{2}
- 0.159\,\delta 
+ 0.028\,\chi_{+}^{4}\right.\\ &\left.
- 0.433\,\chi_{+}\chi_{-}
- 0.426\,\chi_{-}
+ 0.261\,\chi_{+}^{2}
+ 0.890\,\chi_{-}\delta
- 1.006\,\chi_{-}^{2}\delta
+ 0.158\,\chi_{+}^{3}
- 0.546\,\chi_{-}\delta^{3}
- 0.252\,\chi_{+}^{3}\delta\right.\\ &\left.
+ 0.347\,\chi_{+}^{2}\chi_{-}^{2}
- 0.197\,\chi_{+}^{2}\delta
+ 0.206\,\chi_{+}^{2}\chi_{-}
- 0.298\,\chi_{+}^{2}\chi_{-}\delta
\right)\,.
\end{aligned}
\end{equation}
\end{widetext}
\begin{widetext}
\begin{equation}
\begin{aligned}
&p_{33}(\chi_+,\chi_-,\delta)\simeq\;
 -0.414
- 0.162\,\chi_+
+ 0.044\,\chi_+^{3}
+ 1.956\,\chi_- 
+ 0.206\,\chi_-^{2}  0.936\,\chi_+ \chi_-
+ 0.250\,\chi_+\chi_-^{2}\\
&
+ 0.206\,\chi_+^{2}\chi_- + 0.988\,\delta
- 6.288\,\chi_- \delta
+ 0.794\,\chi_+ \delta
- 1.191\,\chi_+\chi_- \delta+ 5.245\,\chi_- \delta^{2}
- 1.203\,\chi_+ \delta^{2}
- 3.172\,\delta^{2}\\
&+ 3.355\,\delta^{3}
- 1.203\,\chi_+ \delta^{3}
\,.
\end{aligned}
\end{equation}
\end{widetext}

For the $(2,1)$ mode we obtain
\begin{widetext}
\begin{equation}
\begin{aligned}
&A_{21}(\delta,\chi_{+},\chi_{-}) \simeq\;
0.2987\,|\chi_{-}|
+\delta\left( 2.6240\,
- 7.5622\,\chi_{-}
+ 11.9443\,\chi_{-}\delta
- 3.4425\,\chi_{+}
- 1.5529\,\delta
+ 6.9491\,\chi_{+}\delta\right.\\ &\left.
- 4.8801\,\chi_{-}\delta^{2}
+ 3.5889\,\chi_{+}\chi_{-}
- 3.4132\,\chi_{+}^{2}\chi_{-}^{2}
- 4.4565\,\chi_{+}\delta^{2}
- 0.5714\,\chi_{+}^{4}
- 1.5029\,\chi_{+}^{2}\chi_{-}
- 3.7079\,\chi_{+}\chi_{-}\delta\right.\\ &\left.
- 0.2406\,\chi_{+}^{3}\delta\right)
\,.
\end{aligned}
\end{equation}

\begin{equation}
\begin{aligned}
&p_{21}(\delta,\chi_{+},\chi_{-}) \simeq\;
 0.320\,\chi_{+}
+ 0.145
+ 0.090\,\chi_{+}^{2}
+ 1.964\,\chi_{-}\delta
- 0.247\,\chi_{+}\chi_{-}^{2}  - 0.809\,\delta^{5}
+ 2.424\,\chi_{+}^{2}\delta^{2}\\
& 
+ 1.442\,\chi_{-}\delta^{4}
- 1.986\,\chi_{-}^{3}
+ 4.333\,\chi_{-}^{3}\delta - 3.180\,\chi_{+}^{2}\delta^{3}
- 1.974\,\chi_{-}^{2}\delta^{3}
- 1.573\,\chi_{-}^{4}
- 5.663\,\chi_{-}\delta^{3}  + 2.433\,\chi_{-}^{4}\delta\\
&
- 0.064\,\chi_{-}^{2}
+ 0.497\,\chi_{+}^{3}
- 0.263\,\chi_{+}^{5}  - 0.455\,\chi_{+}^{2}\chi_{-}^{2}\delta
- 0.107\,\delta
+ 1.284\,\chi_{-}^{2}\delta
+ 1.646\,\chi_{+}^{2}\chi_{-}^{3} \\
& - 0.798\,\chi_{+}^{4}\chi_{-}
+ 2.464\,\chi_{+}\chi_{-}^{3}\delta
+ 0.744\,\chi_{+}^{3}\chi_{-}^{2} - 0.750\,\chi_{+}\chi_{-}^{2}\delta^{2}\,.
\end{aligned}
\end{equation}
\end{widetext}
For the $(4,4)$ mode we obtain
\begin{widetext}
\begin{equation}
\begin{aligned}
&A_{44}(\delta,\chi_{+},\chi_{-}) \simeq\;
 0.288
- 0.312\,\chi_{-}^{3}
+ 2.159\,\delta^{4}
+ 0.506\,\chi_{+}^{2}
+ 0.764\,\chi_{+}\chi_{-}- 1.484\,\delta^{2}
+ 0.162\,\chi_{-}
- 0.135\,\chi_{-}^{2}\\
& 
+ 0.030\,\chi_{+}^{4}
+ 0.002\,\delta + 1.334\,\chi_{+}^{2}\delta^{2}
+ 0.921\,\chi_{+}\chi_{-}^{2}
- 1.709\,\chi_{+}^{2}\chi_{-}\delta
+ 0.819\,\chi_{+}^{3} - 0.988\,\chi_{+}\chi_{-}\delta^{2}
+ 0.038\,\chi_{+}\chi_{-}^{2}\delta \\
&
- 0.407\,\chi_{-}\delta^{3}
- 1.562\,\chi_{+}\chi_{-}^{3}- 1.707\,\chi_{+}^{2}\delta
+ 1.896\,\chi_{+}^{2}\chi_{-}^{2}
+ 0.353\,\chi_{-}^{4}
+ 0.571\,\chi_{-}^{3}\delta + 1.538\,\delta^{3}
- 0.784\,\chi_{+}^{3}\delta\\
& - 0.864\,\chi_{+}^{3}\chi_{-}
- 0.492\,\chi_{+}^{2}\chi_{-}  - 0.692\,\chi_{+}\chi_{-}\delta \,.
\end{aligned}
\end{equation}

\begin{equation}
\begin{aligned}
&p_{44}(\delta,\chi_{+},\chi_{-}) \simeq\;
 2.593
- 132.719\,\delta^{3}
- 0.441\,\chi_{-}
- 10.437\,\chi_{+}^{2}\delta
+ 19.225\,\delta  + 116.664\,\delta^{5}
+ 27.486\,\chi_{+}^{2}\delta^{3}\\
&
+ 18.241\,\chi_{+}\delta
- 10.757\,\chi_{+}^{3}  - 14.336\,\chi_{-}\delta
- 15.505\,\chi_{-}^{3}\delta^{2}
- 3.197\,\chi_{-}^{4}
- 11.866\,\chi_{+}\chi_{-}^{2}  + 12.671\,\chi_{+}^{5}
+ 1.488\,\chi_{-}^{3}\\
&
- 21.751\,\chi_{+}\delta^{2}
+ 49.066\,\chi_{-}\delta^{4} + 12.432\,\chi_{+}^{2}\chi_{-}^{3}
+ 25.562\,\chi_{-}^{3}\delta
+ 44.443\,\delta^{4}
+ 0.389\,\chi_{-}^{2}\,.
\end{aligned}
\end{equation}

\begin{figure}[ht]
    \centering
    \includegraphics[width=0.495\textwidth]{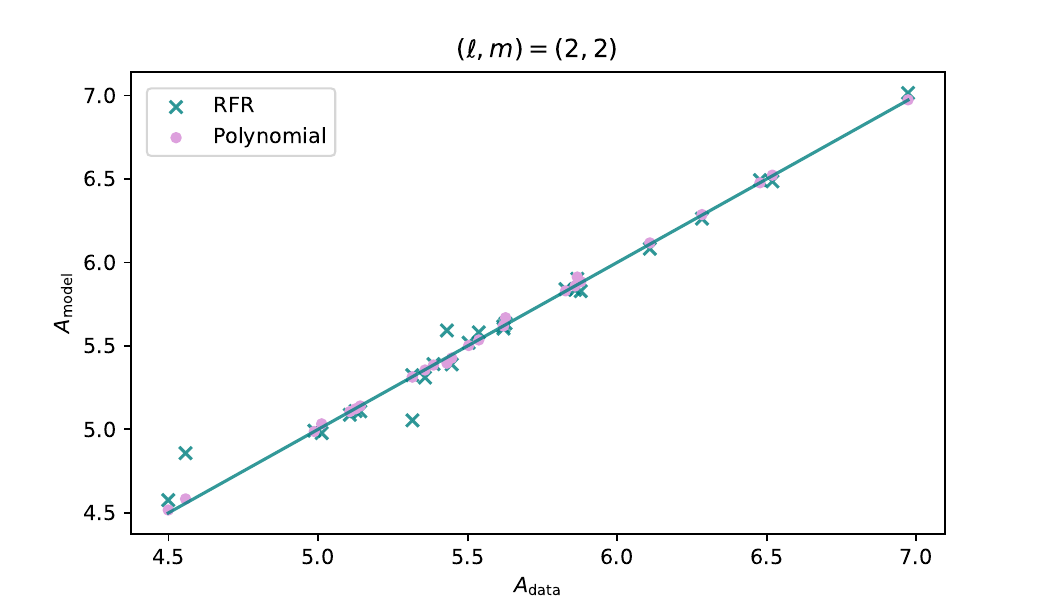}
    \includegraphics[width=0.495\textwidth]{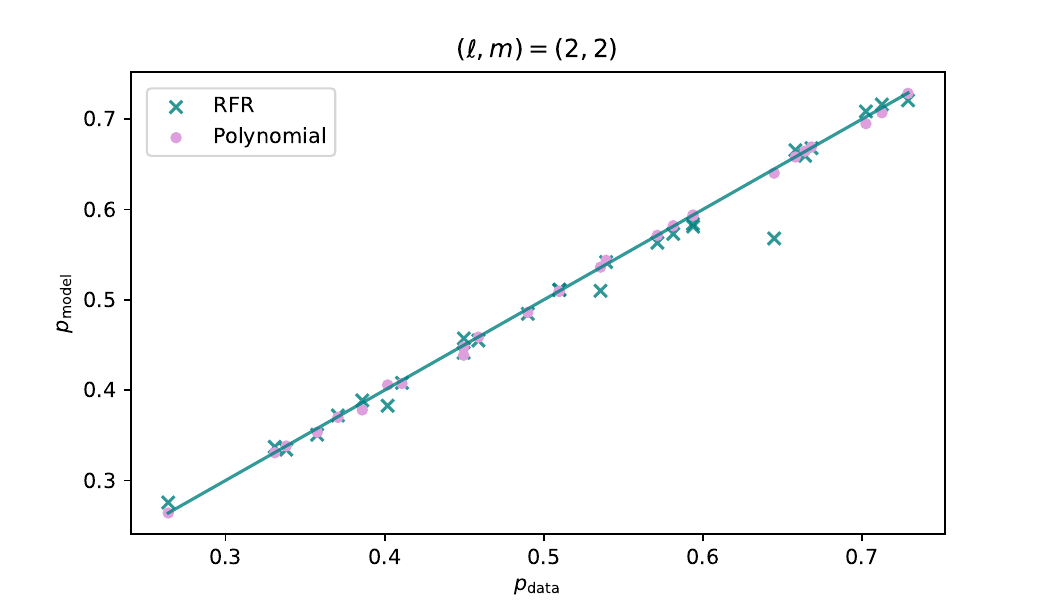}
    \includegraphics[width=0.495\textwidth]{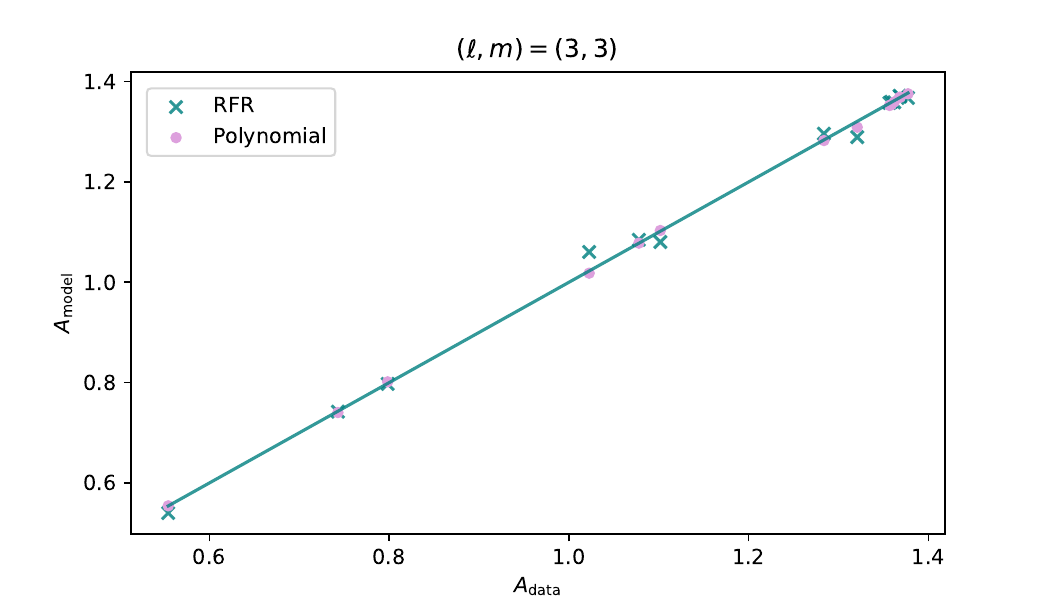}
    \includegraphics[width=0.495\textwidth]{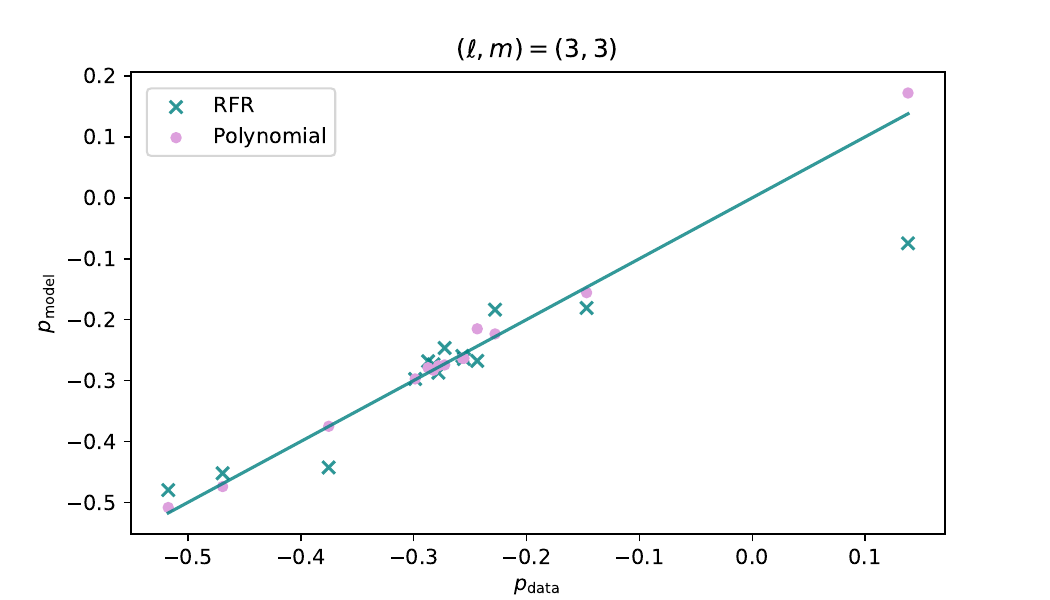}
    \includegraphics[width=0.495\textwidth]{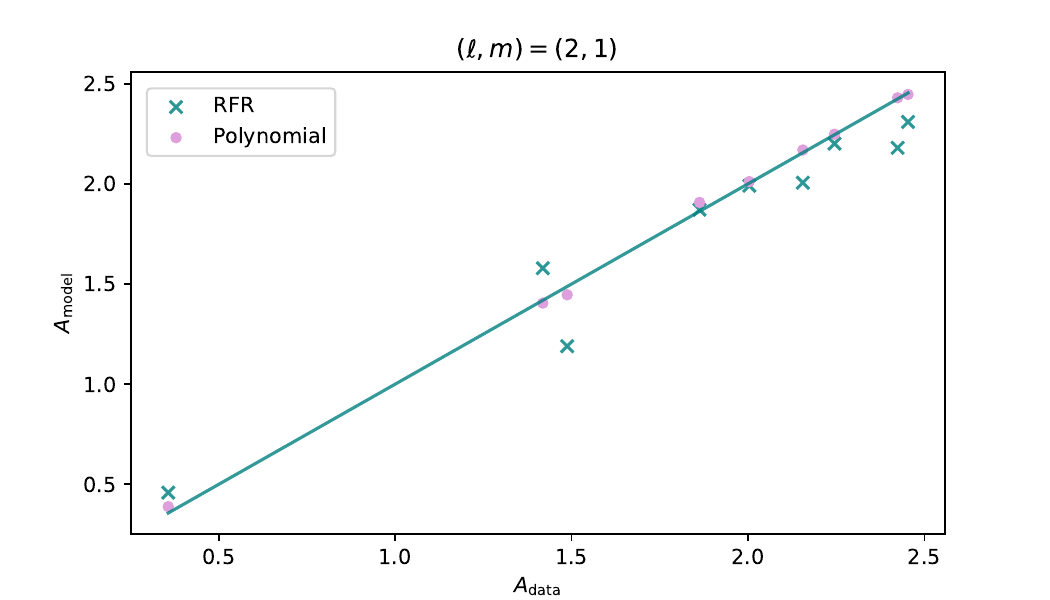}
    \includegraphics[width=0.495\textwidth]{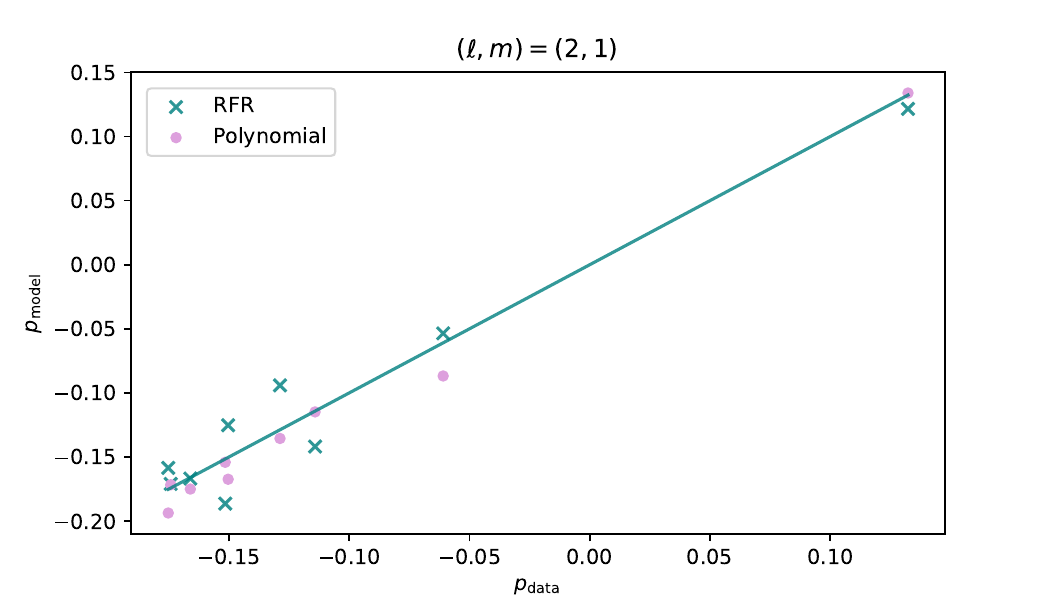}
    \includegraphics[width=0.495\textwidth]{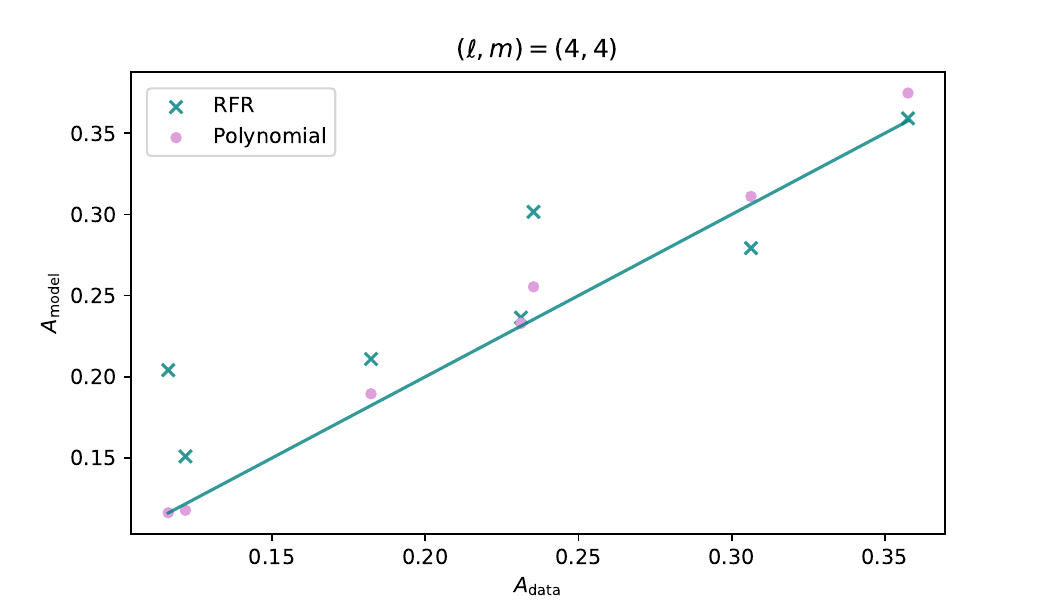}
    \includegraphics[width=0.495\textwidth]{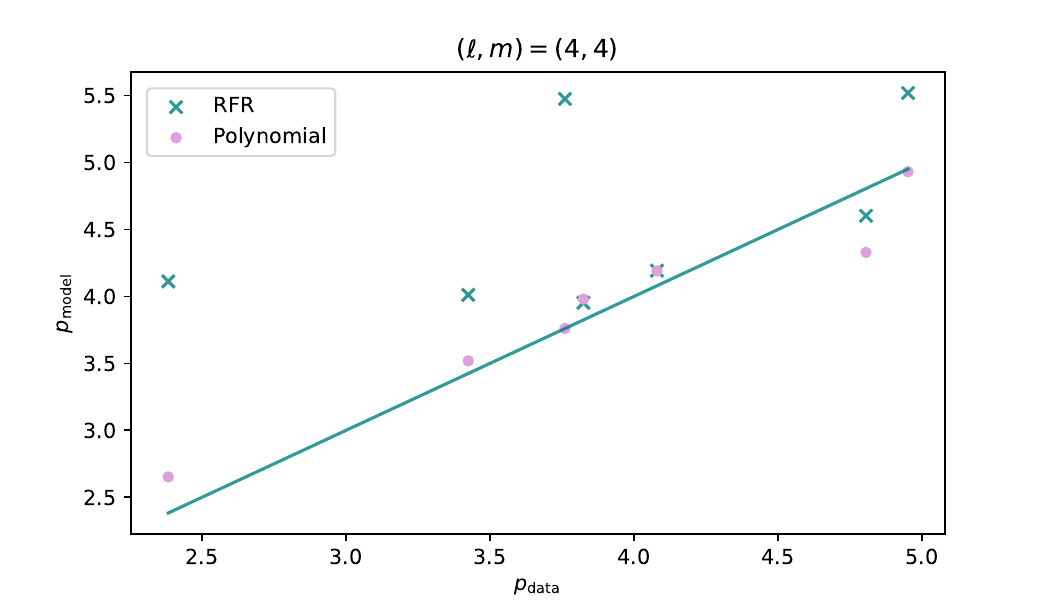}
    \caption{
    Predictions for $A_{\ell m}$ (left) and $p_{\ell m}$ (right) as functions of $[\delta,\,\chi_+,\,\chi_-]$ determined by a RFR (teal) and the polynomial model of Eq.~\eqref{eq:poly_model} (light purple) trained over the same sets of SXS simulations. Results are shown for modes $(\ell, m)=(2,2), (3,3), (2,1)$ and $(4,4)$, going from the top to the bottom rows. Both models generally predict the parameters $A_{\ell m}$ and $p_{\ell m}$ fairly well, with the polynomial fit performing slightly better. Notable exceptions are for $p_{33}$ and $p_{21}$, where the low amplitude of the multipoles when $q\simeq 1$ and the lower number of simulations contribute to worse overall RFR and polynomial fits. The regressor also struggles to fit the $(4,4)$ parameters given the much smaller training set.
    }
    \label{fig:rfr_vs_poly}
\end{figure}
\end{widetext}

In Fig.~\ref{fig:rfr_vs_poly}, we show how the predictions of these polynomial fits compare with the predictions of the RFR, for which we use the \texttt{RandomForestRegressor} module available from \texttt{scikit-learn}~\cite{scikit-learn}. We train the RFR over all our simulations with $\eta>0.15$ (this corresponds to $q<4.4$), due to the relatively low number of simulations at lower $\eta$ (higher $q$). For odd-$m$ modes, we further restrict the training to simulations with $q>1$, as the mode amplitudes for the $q=1$ simulations are too low to obtain reliable FFTs. The results in Fig.~\ref{fig:rfr_vs_poly} show the best RFR predictions optimized over a number of different hyperparameters (number  of estimators, maximum depth, etc.).
From Fig.~\ref{fig:rfr_vs_poly}, it is clear that the polynomial fit performs at least as well, if not somewhat better, than the nonparametric RFR model. We take this to be an indication of the goodness of our polynomial models.

\subsection*{Assessing the quality and applicability of the fits} 

From Fig.~\ref{fig:rfr_vs_poly}, it is
immediately clear that the most accurate results are obtained for the dominant
$(2,2)$ mode.  It is therefore natural to ask in which region of the parameter
space the $(2,2)$ hyper-fit attains its highest reliability. 

\begin{figure}[ht]
    \centering
    \includegraphics[width=0.8\linewidth]{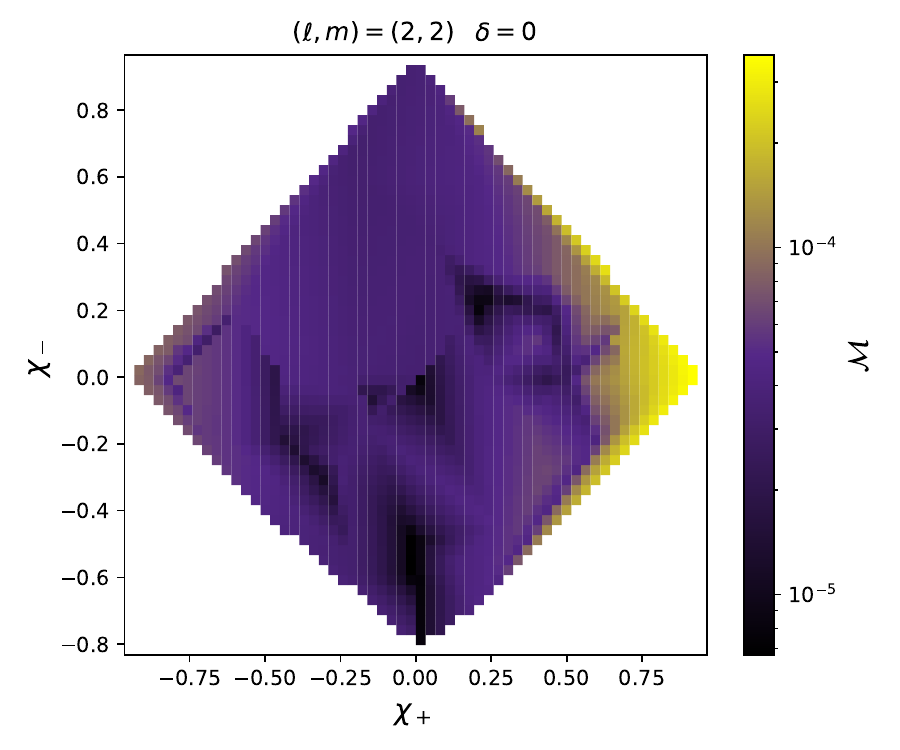} \includegraphics[width=0.8\linewidth]{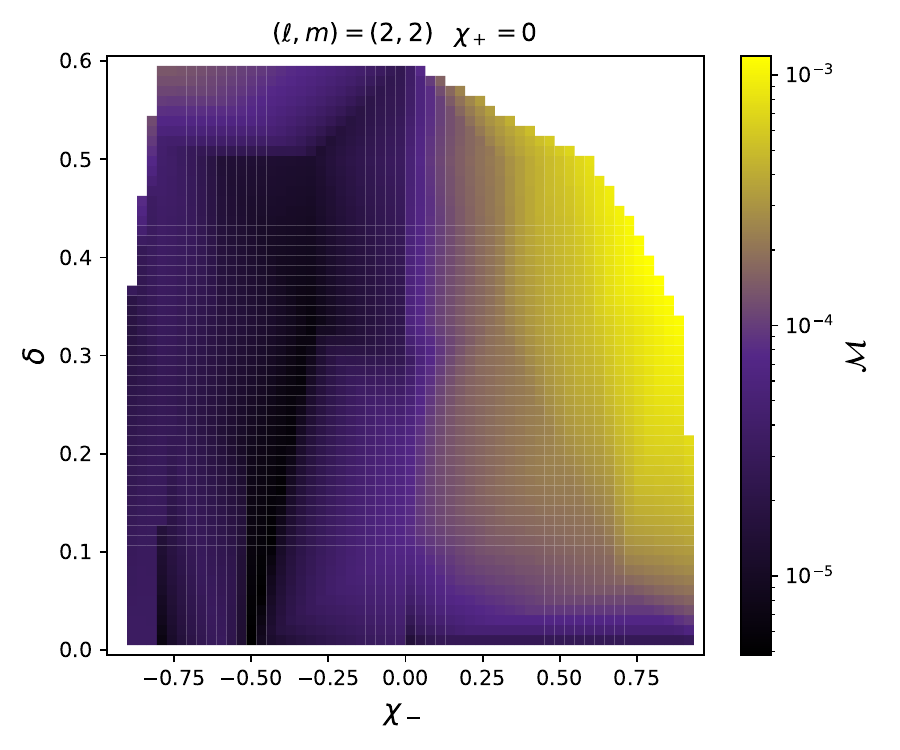} \includegraphics[width=0.8\linewidth]{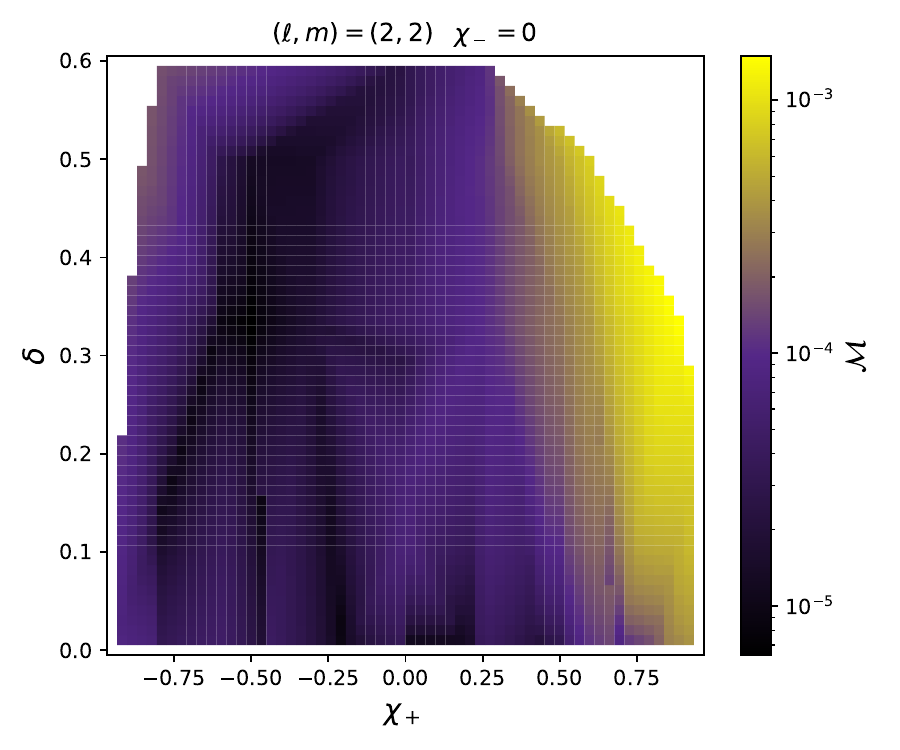}
    \caption{%
    Mean mismatch for the $(2,2)$ mode across different slices of the parameter
    space.  
    Top panel: slice $\delta\simeq 0$.  
    Middle panel: slice $\chi_+\simeq 0$.  
    Bottom panel: slice $\chi_-\simeq 0$.  
    The color scale highlights the regions where the hyper-fit of 
    $(A_{22},p_{22})$ yields the largest deviations from the SXS spectra.  
    The mismatch remains below ${\cal O}(10^{-3})$ everywhere, but the accuracy
    deteriorates noticeably near the edges of the parameter space, in particular
    for large values of $\chi_+$ and $\chi_-$.%
    }
    \label{fig:colormapmismatch}
\end{figure}

To address this question, in Fig.~\ref{fig:colormapmismatch} we show the mean
mismatch over three representative slices of the three-dimensional parameter
space $(\delta,\chi_+,\chi_-)$, namely:
(i) the slice $\delta=0$,  
(ii) the slice $\chi_+=0$, and  
(iii) the slice $\chi_-=0$,  
each selected with a tolerance of $2\%$.  
This allows us to visualize how the performance of the model varies across the
parameter space while retaining good sampling in each direction.

As shown in Fig.~\ref{fig:colormapmismatch}, the mismatch is small in the bulk of the parameter space and increases only near its boundaries, where the available waveforms are more sparse and the spectral shape becomes more extreme (typically at large $\chi_+$ and $\chi_-$). However, 
even in these less favorable regions, the mismatch remains at most $\mathcal{O}(10^{-3})$, while for typical configurations it is of order $\mathcal{O}(10^{-4})$ or smaller.  
This confirms the robustness of the hyper-fit representation for $(A_{22},p_{22})$ across the entire physically relevant domain.

\subsection{How to apply the hyper-fits}\label{app:hyperfitapplication}

When applying the model to numerical data, it is important to select the frequency window in which the power-law behavior of the spectrum is reliably captured and the data are not compromised by the noise.  
As clearly visible in Fig.~\ref{fig:fit_quality} (and consistently across all simulations), the magnitude of the Fourier transform $H_{\ell m}(\omega)$ exhibits two regimes: a power-law decay at low frequencies followed by an exponential decay at high frequencies.  
In logarithmic scale, the transition between the power-law regime and the exponential decay occurs near the point of maximum downward concavity of the spectrum, i.e., the frequency at which the curvature attains its most negative value.  
We denote this frequency by $\bar{\omega}_*$ and determine it independently for each simulation.  
We denote by $\bar{\omega}_*$ the frequency at which this transition occurs. This value can be estimated independently for each simulation.

The hyper-fit model of Eq.~\eqref{eq:model} is designed to be applied in the frequency interval
\[
\bar{\omega} \in [\bar{\omega}_{\rm in},\,\bar{\omega}_{\rm f}]\,,
\]
where the initial point $\omega_{\rm in}$ is chosen relative to the knee
frequency $\bar{\omega}_*$.  
On the basis of our systematic analysis, we recommend the following criteria:
\begin{enumerate}
    \item A safe, universal choice is
\[
\bar{\omega}_{\rm in} \gtrsim 0.75\,\bar{\omega}_*\,,
\]
which ensures that the model is applied well within the power-law regime for
the vast majority of simulations.

\item The upper frequency $\bar{\omega}_{\rm f}$ can be defined as the frequency at which
the magnitude of the Fourier transform decreases to one twentieth of its value
at $\bar{\omega}_*$.  
This prescription safely excludes the region where numerical noise affects the
Fourier transform across the entire catalog.  
However, for simulations in which the Fourier-transform amplitude is
intrinsically very small, the numerical noise floor may be reached earlier. In
such cases, the value of $\bar{\omega}_{\rm f}$ may need to be lowered accordingly to
ensure a reliable fit.

\end{enumerate}

These prescriptions guarantee that the fitted parameters $(A_{\ell m},p_{\ell m})$
are evaluated in a region where the spectral model is both stable and
physically meaningful. When these conditions are satisfied, the hyper-fits provide an accurate description of the data.

\section{Detector-Weighted Mismatch Analysis}\label{app:AdvLIGO}
In this section we assess the agreement between the numerical data and the parametric model by means of a detector-weighted mismatch. The mismatch is computed according to the definition given in Eq.~\eqref{eq:mismatch}, where the inner product is evaluated by weighting the frequency-domain spectra with the one-sided noise power spectral density of Advanced LIGO~\cite{LIGOsensitivitycurve}. We assume a remnant BH mass of \(M = 60\,M_\odot\). The noise-weighted inner product is defined as
\begin{equation}\label{eq:sensitivity}
\langle a \mid b \rangle
=  \int_{\bar{\omega}_{\rm i}}^{\bar{\omega}_{\rm f}}
\frac{\tilde a(\bar{\omega})\,\tilde b^{*}(\bar{\omega})}{S_n(\bar{\omega})}\,\mathrm{d}\bar{\omega} \,,
\end{equation}
where \(S_n(f)\) denotes the one-sided noise power spectral density.
\begin{figure}[H]
    \centering    \includegraphics[width=\linewidth]{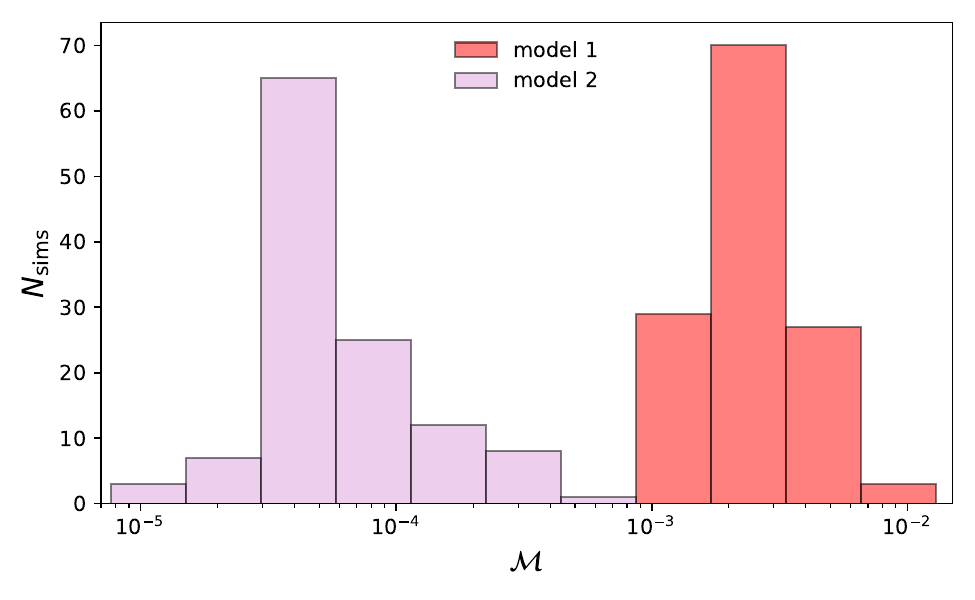}
    \caption{%
    Distribution of detector-weighted mismatches between the numerical SXS $(2,2)$ modes and the two models discussed in Sec.~\ref{sec:modelcomparison}, computed using the Advanced LIGO sensitivity curve. Model $1$ corresponds to $H^{(1)}=MA_{22}\,\sqrt{\mathcal{R}_{22}}$, while model $2$ corresponds to $H^{(2)}=MA_{22}\,\sqrt{\mathcal{R}_{22}}/\bar{\omega}^{p_{22}}$.  
    We consider the same set of simulations of Fig.~\ref{fig:mismatch_histogram}.  
    For each simulation, the mismatch is computed using Eq.~\eqref{eq:mismatch} with the noise-weighted inner product defined in Eq.~\eqref{eq:sensitivity}.}
    \label{fig:mismatch_histogram_AdvLIGO}
\end{figure}

The resulting distribution of mismatches using the Advanced LIGO sensitivity curve is shown in Fig.~\ref{fig:mismatch_histogram_AdvLIGO} and it is nearly indistinguishable from Fig.~\ref{fig:mismatch_histogram}, indicating that the inclusion of detector noise does not alter the relative performance of the model for a GW150914-like binary. 
The model $H^{(2)}$ of Eq.~\eqref{eq:model}, with parameters $(A_{22},p_{22})$ obtained from the hyper-fit formulas of Appendix~\ref{app:fitresults}, yields mismatches that remain again below $10^{-3}$ in all cases, and typically at the level of $\mathcal{O}(10^{-5}-10^{-4})$.  
In contrast, the simpler model $H^{(1)}$ produces a distribution shifted toward larger values, with mismatches generally one to two orders of magnitude higher.  
These results confirm the superior accuracy of the $H^{(2)}$ model in reproducing the numerical SXS $(2,2)$ data across the parameter space, even when detector sensitivity is explicitly taken into account.  
The level of mismatch can be directly compared with that reported for QNM-based post-merger models (see Fig.~3 of Ref.~\cite{Crescimbeni:2025ytx}).
This demonstrates that the model has good mismatches under realistic assumptions on the detector sensitivity for current ground-based detectors.


\bibliography{ringdown_gf}

\end{document}